\documentclass[twocolumn,showpacs,superscriptaddress,aps,prb]{revtex4-1}
\usepackage{bm}
\usepackage{amssymb,amsmath}

\usepackage[dvips]{graphicx}

\begin{document}

\title{Emerging criticality in the disordered three-color Ashkin-Teller model}

\author{Qiong Zhu}
\affiliation{Zhejiang Institute of Modern Physics, Zhejiang University, Hangzhou 310027, China}

\author{Xin Wan}
\affiliation{Zhejiang Institute of Modern Physics, Zhejiang University, Hangzhou 310027, China}
\affiliation{Collaborative Innovation Center of Advanced Microstructures, Nanjing 210093, China}

\author{Rajesh Narayanan}
\affiliation{Department of Physics, Indian Institute of Technology Madras, Chennai 600036, India}

\author{Jos\'e A. Hoyos}
\affiliation{Instituto de F\'{\i}sica de S\~ao Carlos, Universidade de S\~ao Paulo,
C.P. 369, S\~ao Carlos, S\~ao Paulo 13560-970, Brazil}

\author{Thomas Vojta}
\affiliation{Department of Physics, Missouri University of Science and Technology, Rolla, Missouri 65409, USA}

\begin{abstract}
We study the effects of quenched disorder on the first-order phase transition in the
two-dimensional three-color Ashkin-Teller model by means of large-scale Monte Carlo
simulations. We demonstrate that the first-order phase transition is rounded by the
disorder and turns into a continuous one. Using a careful finite-size-scaling analysis,
we provide strong evidence for the emerging critical behavior of the disordered Ashkin-Teller
model to be in the clean
two-dimensional Ising universality class, accompanied by universal logarithmic
corrections. This agrees with perturbative renormalization-group predictions by Cardy.
As a byproduct, we also provide support for the strong-universality scenario
for the critical behavior of the two-dimensional disordered Ising model.
We discuss consequences of our results for the classification of disordered phase transitions
as well as generalizations to other systems.
\end{abstract}

\date{\today}
\pacs{75.10.Nr, 75.40.-s, 05.70.Jk}

\maketitle
%%%%%%%%%%%%%%%%%%%%%%%%%%%%%%%%%%%%%%%%%%%%%%%%%%%%%%%%%%%%%%%%%%%%%%%%%%%%%%%%%
% Main text starts here
%%%%%%%%%%%%%%%%%%%%%%%%%%%%%%%%%%%%%%%%%%%%%%%%%%%%%%%%%%%%%%%%%%%%%%%%%%%%%%%%%
\section{Introduction}
\label{Intro}
%%%%%%%%%%%%%%%%%%%%%%%%%%%%%%%%%%%%%%%%%%%%%%%%%%%%%%%%%%%%%%%%%%%%%%%%%%%%%%%%%

The Imry-Ma criterion  \cite{ImryMa75,*ImryWortis79,*HuiBerker89} is one of the key results on phase transitions in disordered systems.
It governs the stability of macroscopic phase coexistence against quenched random disorder that
locally favors one phase over the other. By comparing the energy gain due to the disorder
with the energy cost of a domain wall, Imry and Ma showed that disorder destroys phase
coexistence by domain formation in dimensions $d\le 2$. \footnote{If the randomness breaks a continuous symmetry, the marginal
dimension is $d = 4$.}
As a consequence, infinitesimal disorder rounds first-order phase transitions in
$d\le 2$ as Aizenman and Wehr \cite{AizenmanWehr89} later proved rigorously as a theorem.
\footnote{The question of whether or not phase coexistence can survive in three dimensions attracted
a lot of attention in the context of the random-field Ising model. It was answered affirmatively in Ref.\
\onlinecite{Imbrie84,BricmontKupiainen87}.}

These results raise the important question of what is the fate of a first-order transition
that is destroyed by disorder? Is it a continuous
transition, does an intermediate phase appear, or is the sharp transition, perhaps,
completely destroyed via smearing? If the transition becomes continuous, what is the critical behavior?
Is it accompanied by pretransitional singularities due to rare regions, as is the case at
``generic'' critical points in disordered systems (see, e.g.,
Refs.\ \onlinecite{Vojta06,Vojta10})? These questions have recently reattracted considerable attention,
in particular in the context of zero-temperature quantum phase transitions.\cite{SenthilMajumdar96,
GoswamiSchwabChakravarty08,GreenblattAizenmanLebowitz09,HrahshehHoyosVojta12,Barghathietal14}

It turns out, however, that these questions remain unresolved even for a simple prototypical classical
phase transition, viz., the transition in the two-dimensional ferromagnetic Ashkin-Teller model. The $N$-color
Ashkin-Teller model \cite{AshkinTeller43,GrestWidom81,Fradkin84,Shankar85} consists of $N$ Ising models,
coupled via their energy densities. In the absence of disorder and for $N>2$, this system features a first-order
phase transition between a paramagnetic high-temperature phase and a ferromagnetic (Baxter) phase
a low temperatures. According to the Imry-Ma criterion, or equivalently the Aizenman-Wehr theorem, this first-order transition cannot survive
the introduction of weak disorder in the form of random bonds or bond or site dilution. Murthy \cite{Murthy87} and
Cardy \cite{Cardy96,*Cardy99} analyzed this problem by means of perturbative renormalization group calculations
which predicted that the first-order transition is rounded to a continuous transition in the universality class
of the two-dimensional \emph{clean} Ising model, apart from logarithmic corrections. However, recent numerical simulations
of a random-bond three-color Ashkin-Teller model
\cite{BellafardKatzgraberTroyerChakravarty12,BellafardChakravartyTroyerKatzgraber14} disagreed with
these predictions. They found nonuniversal critical exponents that vary with disorder strength and
differ from the clean Ising exponents. Moreover, the reported value of the correlation length exponent $\nu$ violates
the inequality $d\nu \ge 2$ due to Chayes et al.\cite{CCFS86}

To resolve these contradicting results, we perform large-scale high-accuracy Monte Carlo simulations
of the two-dimensional three-color Ashkin-Teller model. We consider two types of quenched disorder, random
bonds as well as site dilution. Our data provide strong evidence that the emerging critical behavior is universal
and in the clean Ising universality class, as predicted by the renormalization group calculations.\cite{Murthy87,Cardy96,*Cardy99}
It is also accompanied by logarithmic corrections analogous to those found in the disordered two-dimensional Ising model.

The rest of our paper is organized as follows:  In Sec.\ \ref{Model}, we introduce the $N$-color Ashkin-Teller model
and discuss its properties in the absence of disorder. We then briefly summarize the results of Cardy's renormalization
group theory. In Sec.\ \ref{MC}, we explain our Monte Carlo method, and we give an overview over the simulation
parameters. Section \ref{Results} is devoted to the numerical results for the clean, site-diluted, and random-bond
Ashkin-Teller models. As a byproduct, our data provide additional support for the strong-universality scenario
for the two-dimensional disordered Ising model.
We conclude in Sec.\ \ref{Conclusions} by discussing consequences of our results for the
classification of disordered phase transitions as well as generalizations to other systems.

%%%%%%%%%%%%%%%%%%%%%%%%%%%%%%%%%%%%%%%%%%%%%%%%%%%%%%%%%%%%%%%%%%%%%%%%%%%%%%%%%
\section{Model and theory}
\label{Model}
%%%%%%%%%%%%%%%%%%%%%%%%%%%%%%%%%%%%%%%%%%%%%%%%%%%%%%%%%%%%%%%%%%%%%%%%%%%%%%%%%

The two-dimensional $N$-color Ashkin-Teller model \cite{GrestWidom81,Fradkin84,Shankar85} is a generalization of the original model
proposed by Ashkin and Teller \cite{AshkinTeller43} (which corresponds to the $N=2$ case). It consists of $N$ identical
Ising models, coupled via their energy densities. The Hamiltonian of the clean model reads
\begin{equation}
H = -J \sum_{\alpha=1}^N \sum_{\langle ij \rangle} S_i^{\alpha} S_j^{\alpha}
    -K \sum_{\alpha<\beta} \sum_{\langle ij \rangle} S_i^{\alpha} S_j^{\alpha} S_i^{\beta} S_j^{\beta}~.
\label{eq:HAT}
\end{equation}
Here, $i$ and $j$ denote the sites of a regular square lattice of $L^2$ sites, and the corresponding sum is over pairs of nearest neighbors.
$\alpha$ is the ``color'' index that distinguishes the $N$ Ising models, and $S_i^\alpha = \pm 1$ are the usual classical
Ising variables. We are interested in the regime in which both the Ising interaction $J$ and the four-spin interaction $K$
are positive. The strength of the coupling between the Ising models can be parameterized by the dimensionless ratio
$\epsilon=K/J$.
Note that the Hamiltonian (\ref{eq:HAT}) is self dual for the case of $N=2$ colors; and this property
has been used to find the exact location of the phase transition in the clean and disordered models. \cite{WuLin74,WisemanDomany95}
For $N>2$, the Hamiltonian (\ref{eq:HAT}) is not self-dual.  Self-duality can be restored, however, by including higher-order
terms with up to $2N$ spins.\cite{DJLP99} We have not done this in our work, mainly to keep our results quantitatively
comparable to other simulations \cite{GrestWidom81,BellafardKatzgraberTroyerChakravarty12,BellafardChakravartyTroyerKatzgraber14}
in the literature.

The properties of the clean Ashkin-Teller model have been studied in great detail. The two-color model ($N=2$) features a
continuous transition with nonuniversal, continuously varying exponents between a paramagnetic high-temperature phase and
an ordered phase at low temperatures (see, e.g., Ref.\ \onlinecite{Baxter_book82}
 and references therein). In contrast, this transition is of first order for $N>2$, which is the case we are
interested in.\cite{GrestWidom81,Fradkin84,Shankar85}

Quenched disorder can be introduced into the Hamiltonian (\ref{eq:HAT}) in several ways. We consider both site dilution and
bond randomness. In the former case, a fraction $p$ of the lattice sites is removed at random (the $S_i^\alpha$ for \emph{all} colors
$\alpha$ are removed at such vacancy sites). The interactions between the remaining sites retain their uniform values $J$ and $K$.
In the case of bond randomness, the Ising couplings $J_{ij}$ between neighboring sites $i$ and $j$
become independent random variables drawn from some probability distribution $P(J)$ which we take to be a binary distribution
\begin{equation}
P(J) = c\delta(J-J_h) + (1-c) \delta(J-J_l)
\label{eq:P(J)}
\end{equation}
with $J_h > J_l>0$. Here, $c$ is the concentration of the stronger bonds. The four-spin couplings $K_{ij}$ are either taken to be
uniform or they are slaved to the Ising interactions on the same bond via $K_{ij} = \epsilon J_{ij}$ with constant $\epsilon$.
Both site dilution and random bonds are realizations of random-$T_c$ disorder, i.e., disorder that does not break any of the spin symmetries
but changes the local tendency towards the high-temperature or low-temperature phases. Thus, if the system undergoes a
continuous phase transition, both types of disorder should lead to the same universality class.

\begin{figure}
\includegraphics[width=8cm]{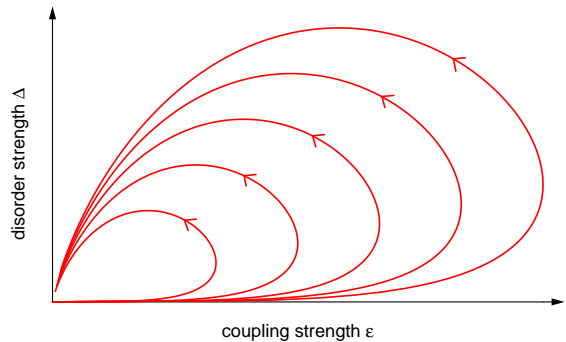}
\caption{(Color online) Cardy's renormalization group trajectories on the critical surface (coupling strength $\epsilon$ vs.\
     disorder strength $\Delta$) for $N=3$.
     The trajectories initially flow towards the first-order region at strong coupling. However, they eventually curl back towards
     the clean Ising fixed point at $\epsilon=\Delta=0$.}
\label{fig:Cardy_flow}
\end{figure}

Murthy \cite{Murthy87} and Cardy \cite{Cardy96,Cardy99} applied a perturbative renormalization group to a continuum version of the two-dimensional
$N$-color Ashkin-Teller model. This analysis benefits from the fact that the first-order phase transition in the clean
model is fluctuation-driven. As a result, the renormalization group is controlled in the limit of small inter-color coupling
and weak disorder. Cardy found the renormalization group trajectories on the critical surface in closed form. In terms of
the coupling strength $\epsilon$ and the dimensionless disorder strength $\Delta$, they read
\begin{equation}
\epsilon = \textrm{const} \times (\Delta/\epsilon)^{(N-2)/N} \exp(-2\Delta/N\epsilon)
\label{eq:Cardy_flow}
\end{equation}
A few characteristic trajectories are shown in Fig.\ \ref{fig:Cardy_flow}.
For weak bare (initial) disorder, the trajectories first run towards the strong-coupling region $\epsilon \gg 1$ where the transition
would turn first order. However, they eventually turn around and curl back towards the clean Ising fixed point at $\epsilon=\Delta=0$.
This not only implies that the transitions has become continuous, in agreement with the Imry-Ma criterion, it also means that the
critical behavior is in the clean Ising universality class.
A more detailed analysis of the renormalization group equations produces
additional logarithmic corrections to the
leading Ising power laws, similar to those found in renormalization group approaches to the disordered Ising model.
\cite{DotsenkoDotsenko83,Shalaev84,Shankar87,*Shankar88,Ludwig88}

Furthermore, the large excursions of the renormalization group trajectories for small bare disorder strength imply a very slow crossover
from the first-order transition of the clean Ashkin-Teller model
to the critical point of the disordered system. This crossover is especially interesting because
$d=2$ is the marginal dimensionality for the Aizenman-Wehr theorem. (First-order transitions are destroyed
by randomness for $d\le 2$ while they can survive for $d>2$.) If the clean system has a strong first-order
transition, the breakup length $L_b$,
beyond which randomness becomes important increases very rapidly with decreasing disorder
strength $\Delta$. In fact, for weak disorder, it is expected\cite{Binder83,GrinsteinMa83}
to follow the exponential $L_b \sim \exp(\textrm{const}/\Delta^{2})$.
This implies that enormous system sizes are necessary to reach the asymptotic regime if the
clean first-order transition is strong and the disorder is weak.

%%%%%%%%%%%%%%%%%%%%%%%%%%%%%%%%%%%%%%%%%%%%%%%%%%%%%%%%%%%%%%%%%%%%%%%%%%%%%%%%%
\section{Monte Carlo simulations}
\label{MC}
%%%%%%%%%%%%%%%%%%%%%%%%%%%%%%%%%%%%%%%%%%%%%%%%%%%%%%%%%%%%%%%%%%%%%%%%%%%%%%%%%
\subsection{Overview}
\label{subsec:Overview}
%%%%%%%%%%%%%%%%%%%%%%%%%%%%%%%%%%%%%%%%%%%%%%%%%%%%%%%%%%%%%%%%%%%%%%%%%%%%%%%%%

To resolve the discrepancy between the renormalization group predictions outlined above and the recent numerical results
of Refs.\ \onlinecite{BellafardKatzgraberTroyerChakravarty12,BellafardChakravartyTroyerKatzgraber14}, we perform
large-scale high-accuracy Monte Carlo simulations of two-dimensional three-color Ashkin-Teller models with site dilution
and/or bond randomness.

As we are interested in the critical behavior, a cluster algorithm is required to reduce the critical slowing down
close to the phase transition. We employ a Wolff embedding algorithm similar to that used by Wiseman and Domany \cite{WisemanDomany95}
for the two-color Ashkin-Teller model. Its basic idea is simple. Imagine fixing all $S_i^{(2)}$ and $S_i^{(3)}$ spins. Then,
the Hamiltonian (\ref{eq:HAT}) is equivalent to an (embedded) Ising model for the $S_i^{(1)}$ spins with effective interactions
\begin{equation}
J_{ij}^\textrm{eff} = J + \epsilon J\left( S_i^{(2)}S_j^{(2)} + S_i^{(3)}S_j^{(3)}\right)~.
\label{eq:Jeff}
\end{equation}
Simulating this Ising model using any valid Monte Carlo algorithm establishes detailed balance
between all states with the same fixed  $S_i^{(2)}$ and $S_i^{(3)}$.
We can construct and simulate analogous embedded Ising models to update $S_i^{(2)}$ and $S_i^{(3)}$. By combing
Monte Carlo updates for all three embedded Ising models we arrive at a valid algorithm
(fulfilling ergodicity and detailed balance between all states) for the entire Hamiltonian (\ref{eq:HAT}).

To simulate the embedded Ising models, we use the efficient Wolff and Swendsen-Wang
cluster algorithms.\cite{Wolff89,SwendsenWang87}
They are only valid if all interactions are ferromagnetic, i.e., if all $J_{ij}^\textrm{eff} \ge 0$. This is fulfilled as long as the coupling strength
$|\epsilon| \le 1/(N-1)$. In our case of three colors, $\epsilon$ therefore must not exceed $1/2$.

Finding the averages, variances, and distributions of observables in disordered systems requires
the simulation of many samples with different disorder configurations. For optimal performance,
one must therefore carefully choose the number $n_s$ of samples (i.e., disorder configurations) and the number
$n_m$ of measurements during the simulation of each sample.\cite{BFMM98,BFMM98b,VojtaSknepnek06}
Assuming statistical independence between
measurements (quite possible with a cluster algorithm), the total variance $\sigma_t^2$
of a particular observable (thermodynamically and disorder averaged)
can be estimated as
\begin{equation}
\sigma_t^2 = (\sigma_s^2 + \sigma_m^2/n_m)/n_s
\end{equation}
where $\sigma_s^2$ is the disorder-induced variance between samples and $\sigma_m^2$ is the
variance of measurements within each sample. As the numerical effort is roughly
proportional to $n_m \, n_s$ (neglecting equilibration for the moment), it is clear
that the best value of $n_m$ is quite small. One might even be tempted to measure only
once per sample. However, with too few measurements, the majority of the computer time
would be spent on equilibration.
These requirements can be balanced by using large numbers $n_s$ of disorder
configurations (ranging from several 10000 to several million in our case) and rather
short runs with a few hundred Monte Carlo measurements per sample. Note that such short runs
lead to biases in several observables, at least if the usual estimators are employed. These
biases can be corrected by improved estimators as is discussed in Appendix
\ref{sec:bias}.

Based on these ideas, we develop two independent Monte Carlo codes, one (referred to as code A) mainly employed for
simulating the site-diluted Ashkin-Teller model, and the other one (code B) used for the random-bond case.

%%%%%%%%%%%%%%%%%%%%%%%%%%%%%%%%%%%%%%%%%%%%%%%%%%%%%%%%%%%%%%%%%%%%%%%%%%%%%%%%%
\subsection{Site-diluted simulations}
\label{subsec:MC_site_diluted}
%%%%%%%%%%%%%%%%%%%%%%%%%%%%%%%%%%%%%%%%%%%%%%%%%%%%%%%%%%%%%%%%%%%%%%%%%%%%%%%%%

All site-diluted simulations use code A. We study impurity concentrations $p=0$ (the clean case), 0.05, 0.1, 0.2, and 0.3. For comparison,
the lattice percolation threshold is at $p_c=0.407253$. The Ising interaction $J$ is fixed
at unity while the coupling strength $\epsilon=K/J$ takes values 0 (the Ising limit), 0.05,
0.1, 0.2, 0.3, and 0.5. The system is tuned through the transition by changing the
temperature $T$. Lattice sizes range from $25^2$ sites to $1600^2$ sites ($2240^2$ sites
for the Ising case, $\epsilon=0$) with periodic boundary conditions.
Data are averaged over up to 4 million disorder configurations for the smaller
systems and over up to 500,000 configurations for the largest ones.
This leads to small statistical errors of the data.

For site-diluted systems, we combine the Wolff single-cluster updates with Swendsen-Wang
multi-cluster updates to equilibrate small isolated clusters of lattice sites that can occur
for larger dilutions.
Specifically, a full Monte Carlo sweep consists of a Swendsen-Wang sweep (for each color)
followed by a Wolff sweep. (A Wolff sweep is defined as a number of cluster
flips such that the total number of flipped spins per color is equal to the number of sites.)
To verify our codes we have also compared the results to those of conventional Metropolis
single-spin updates.\cite{MRRT53}

To estimate the equilibration times, we compare runs with ``hot start'' (initial spin values
are completely random) and ``cold start'' (initially, all spins $S_i^\alpha=1$). Characteristic
equilibration times range from less than 10 sweeps for linear system size $L=50$ to about 40 sweeps
for system size $L=1600$. In our production runs, we therefore employ equilibration periods
of 60 to 100 sweeps and measurement periods of another 100 to 200 sweeps, with measurements taken after
every sweep. Using these parameters, the results of runs with hot and cold starts agree within
our small statistical errors.

Note that simulations of the \emph{clean} Ashkin-Teller model close to its strong
first-order phase transition require longer equilibration times to overcome the supercritical
slowing down associated with first-order transitions. Details will be given in Sec.\ \ref{subsec:Results_clean}.

%%%%%%%%%%%%%%%%%%%%%%%%%%%%%%%%%%%%%%%%%%%%%%%%%%%%%%%%%%%%%%%%%%%%%%%%%%%%%%%%%
\subsection{Random-bond simulations}
\label{subsec:MC_random_bond}
%%%%%%%%%%%%%%%%%%%%%%%%%%%%%%%%%%%%%%%%%%%%%%%%%%%%%%%%%%%%%%%%%%%%%%%%%%%%%%%%%

Using code A, we study the random-bond Ashkin-Teller model with the binary bond distribution
(\ref{eq:P(J)}). The Ising interactions take the values $J_h=2$ or $J_l=0.5$, each with
a probability of 0.5. The four-spin interactions are slaved to the Ising interactions
via $K_{ij}=\epsilon J_{ij}$ with constant coupling strength $\epsilon$. We explore
the cases $\epsilon=0$ (random-bond Ising model), 0.1, 0.2, and 0.5. Lattice sizes range from
$35^2$ to $1120^2$ sites with periodic boundary conditions. The numbers of disorder configurations for each parameter set
range from $10^5$ for the largest systems to $10^6$ for the smallest ones.

Otherwise, the random-bond  simulations are analogous to the site-diluted ones:
Each full Monte Carlo sweep is a combination of a Wolff sweep and a Swendsen-Wang sweep.
Each system is equilibrated from a hot start using 100 full sweeps, the measurement
period is another 100 sweeps.

%DESCRIBE RAJ/QIONG SIMULATIONS

In addition, we use code B to perform simulations of a random-bond Ashkin-Teller model with uniform,
non-random four-spin interaction $K$. Specifically, the Ising interactions take the values $J_h=6/5$ and $J_l=4/5$,
each chosen with a probability of 0.5. This implies that the average Ising interaction is
$J\equiv (J_h+J_l)/2 =1$. The four-spin interactions are non-random and given by $K=\epsilon J$.
Because the effective interactions (\ref{eq:Jeff}) appearing in the embedded Wolff algorithm must be positive
for both values of the Ising interaction, the coupling strength is restricted to $\epsilon \le 2/5$.
In our simulations, $\epsilon$ will be fixed at $0.1$. We simulate lattice sizes ranging
from $24^2$ to $1600^2$ sites with periodic boundary conditions.
The numbers of disorder configurations range from $10^4$ for the largest system to $10^5$ for the smallest one.
Each system is equilibrated from a cold start using 200 full Wolff sweeps, the measurement period
is also 200 Wolff sweeps per temperature; and we take measurements after every four sweeps.

%%%%%%%%%%%%%%%%%%%%%%%%%%%%%%%%%%%%%%%%%%%%%%%%%%%%%%%%%%%%%%%%%%%%%%%%%%%%%%%%%
\subsection{Observables}
\label{subsec:Observables}
%%%%%%%%%%%%%%%%%%%%%%%%%%%%%%%%%%%%%%%%%%%%%%%%%%%%%%%%%%%%%%%%%%%%%%%%%%%%%%%%%

During the simulations, we calculate various thermodynamic quantities such as
the energy $E = [\langle e \rangle]_\textrm{dis}$ and the magnetization $M = [\langle m \rangle]_\textrm{dis}$.
Here $e$ and $m$ stand for individual energy and magnetization measurements, and $\langle \ldots \rangle$ is the
canonical thermodynamic average (which is approximated by the Monte Carlo average over $n_m$ measurements).  The average
$[ \dots ]_\textrm{dis}$ over the disorder distribution is approximated by the average over $n_s$ samples.
Specific heat and magnetic susceptibility are calculated from the fluctuations of $e$ and $m$
as $C = (L^2/T^2) [ \langle e^2 \rangle -\langle e \rangle^2 ]_\textrm{dis}$
and $\chi = (L^2/T) [ \langle m^2 \rangle -\langle m \rangle^2 ]_\textrm{dis}$.
We also measure the product order parameter (or ``polarization'') $M_p=[\langle m_p \rangle]_\textrm{dis}$
with $m_p = (1/L^2) \sum_i  S_i^{\alpha} S_i^{\beta}$ for two different colors $\alpha$ and $\beta$.
The corresponding susceptibility reads $\chi_p = (L^2/T) [ \langle m_p^2 \rangle -\langle m_p \rangle^2 ]_\textrm{dis}$.

Magnetization and susceptibility are averaged over the three colors for increased accuracy, and
all quantities are normalized ``per spin''. Analogously,
the product order parameter and its susceptibility are averaged over the three possible pairs of colors.
The statistical errors of all thermodynamic
quantities are estimated from their fluctuations between disorder configurations.

In addition, we calculate several quantities whose scale dimension
is zero which makes them particularly suitable for a finite-size scaling analysis.
The first such quantity is the Binder cumulant of the magnetization. In a disordered
system, we need to distinguish the average Binder cumulant $g_\textrm{av}$ and its
``global'' counterpart $g_\textrm{gl}$, depending on when the disorder average is performed.  They are defined as
\begin{equation}
g_\textrm{av} = \left[1 - \frac {\langle m^4\rangle}{3\langle m^2\rangle^2} \right]_\textrm{dis}~, \quad
g_\textrm{gl} = 1 - \frac {[\langle m^4\rangle]_\textrm{dis}}{3[\langle m^2\rangle]^2_\textrm{dis}} ~.
\label{eq:Binder}
\end{equation}
The Binder cumulants $g^E_\textrm{av}$ and $g^E_\textrm{gl}$ of the energy can be defined analogously.

The correlation length is calculated via the second moment of the spin-spin correlation function
$G(\mathbf{r}) = (1/L^2) \sum_{i,j,\alpha} \langle S_i^\alpha S_j^\alpha \rangle \delta(\mathbf{r}-\mathbf{r}_{ij})$. \cite{CooperFreedmanPreston82,Kim93,CGGP01}
We again need to distinguish average and ``global'' versions of this quantity, depending on when the disorder average is performed.
They can be obtained efficiently from the Fourier transform $\tilde G(q)$ of the correlation function:
\begin{eqnarray}
\xi_\textrm{av} &=& \left[ \left(\frac{\tilde G(0) -\tilde G (q_\textrm{min})}{q_\textrm{min}^2 \tilde G(q_\textrm{min})} \right)^{1/2} \right]_\textrm{dis}~,
\label{eq:xi_av}\\
\xi_\textrm{gl} &=&  \left(\frac{[\tilde G(0) -\tilde G (q_\textrm{min})]_\textrm{dis}}{q_\textrm{min}^2 [\tilde G(q_\textrm{min})]_\textrm{dis}} \right)^{1/2}~.
\end{eqnarray}
Here, $q_\textrm{min}=2\pi/L$ is the minimum wave number that fits into a system of linear size $L$.

As was mentioned in Sec.\ \ref{subsec:Overview}, short Monte Carlo runs potentially introduce biases into
observables for which a nonlinear operation is performed on the data \emph{before} the disorder
average. In our case, this includes $C$, $\chi$, $g_\textrm{av}$, and
$\xi_\textrm{av}$. As is explained in Appendix \ref{sec:bias}, these biases can be eliminated
by using improved estimators.

To judge the quality of the fits of our data to various mathematical models, we use the reduced weighted
error sum $\bar \chi^2$. For fitting $n$ data points $(x_i,y_i)$ to a function $f(x)$ containing $q$ fit parameters, it is defined as
\begin{equation}
\bar \chi^2 = \frac 1 {n-q} \sum_i \frac{(y_i - f(x_i))^2}{\sigma_i^2}
\end{equation}
where $\sigma_i^2$ is the variance of $y_i$. The fits are of good quality if $\bar \chi^2 \lessapprox 2$.

%%%%%%%%%%%%%%%%%%%%%%%%%%%%%%%%%%%%%%%%%%%%%%%%%%%%%%%%%%%%%%%%%%%%%%%%%%%%%%%%%
\section{Results}
\label{Results}
%%%%%%%%%%%%%%%%%%%%%%%%%%%%%%%%%%%%%%%%%%%%%%%%%%%%%%%%%%%%%%%%%%%%%%%%%%%%%%%%%
\subsection{Clean Ashkin-Teller model}
\label{subsec:Results_clean}
%%%%%%%%%%%%%%%%%%%%%%%%%%%%%%%%%%%%%%%%%%%%%%%%%%%%%%%%%%%%%%%%%%%%%%%%%%%%%%%%%

We first perform a number of simulations (using code A) of the clean model (no dilution, uniform interactions
$J=1, K=\epsilon$)
to test our algorithms and for later comparison with the disordered case.

For $\epsilon=0$, the three-color Ashkin-Teller model is identical to three independent Ising models.
We simulate this model on lattices of $50^2$ to $800^2$ sites, averaging the data over 1000
samples for each size.
The critical temperature is determined, as usual, from the crossing of the magnetic Binder cumulant $g$
for different system sizes $L$. We find $T_c=2.26920(4)$ (the number in brackets indicates the error of the last digit)
in agreement with the exact value $2/\ln(1+2^{1/2})=2.269185\ldots$. Straight power-law fits
(without subleading corrections) of magnetization and susceptibility at $T_c$ as functions of $L$ yield the
critical exponent estimates $\beta/\nu=0.1253(4)$ and $\gamma/\nu=1.751(2)$. The correlation length
exponent itself derives from the temperature derivative of $g$ at criticality. We find
$\nu=0.992(7)$. All fits are of good quality (reduced $\bar \chi^2$ of about $0.3 \ldots 1.2$), and the
estimates are in excellent agreement with the exact values $\beta=1/8, \gamma=7/4$,
and $\nu=1$.

\begin{figure}
\includegraphics[width=8.7cm]{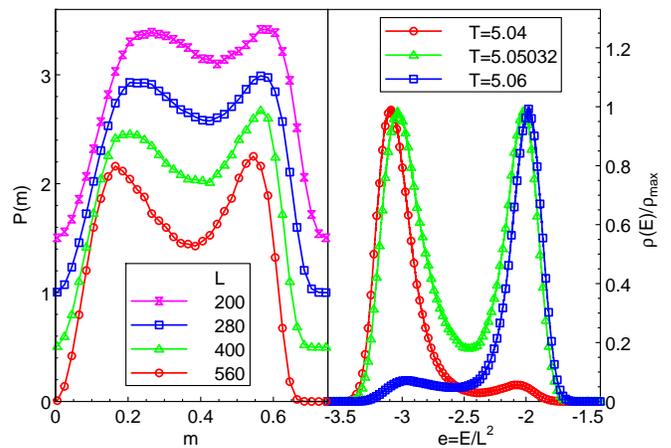}
\caption{(Color online) Left: Magnetization distribution $P(m)$ of the clean three-color Ashkin-Teller model
for coupling $\epsilon=0.3$ close to the transition temperature $T_c \approx 3.178$. The double-peak
structure characteristic of a first-order transition becomes more pronounced with increasing system size.
(The curves for $L=400,280$, and 200 are shifted upwards by multiples of 0.5 for clarity.)
Right: Wang-Landau density of states $\rho(E)$ weighted by the Boltzmann factor (and normalized to its maximum)
for the clean Ashkin-Teller model at $\epsilon=1$. The system size is $L=48$.}
\label{fig:P(m)_clean}
\end{figure}

For nonzero positive $\epsilon$, the phase transition in the clean Ashkin-Teller model is known
\cite{GrestWidom81,Fradkin84,Shankar85} to be of first-order. We confirm this by simulations
for $\epsilon=0.2, 0.3$ and 0.5 using systems of up to $560^2$ sites, averaged over 1000
samples. Because of the supercritical slowing down associated with first-order transitions we
increase the equilibration period to up to 500 Monte Carlo sweeps and the measurement period to 2000
sweeps. The first-order character can be seen clearly in the double-peak structure of the probability
distribution $P(m)$ of the magnetization close to the transition temperature, as shown in Fig.\ \ref{fig:P(m)_clean}.
Notice that the minimum between the peaks becomes more pronounced with increasing system size, and the
distance between the peaks remains roughly unchanged.

We also perform exploratory simulations for $\epsilon=0.7$ and 1.0. The Wolff and Swendsen-Wang
algorithms are invalid for these values because the effective interactions (\ref{eq:Jeff}) can become negative.
We therefore employ only Metropolis updates which requires long equilibration and measurement times
and severely restricts the possible system sizes.
Consequently, our simulations of up to $200^2$ lattice sites
(using 1000 samples, each with 5000 equilibration sweeps and 10000 measurement sweeps) are less accurate
than the simulations for $\epsilon \le 0.5$. However, by comparing runs with ``hot'' and ``cold'' starts
we can bracket the transition temperature with reasonable precision.
Analogous simulations of the clean system are also carried out using code B.

The supercritical slowing down can be overcome
by alternative sampling approaches. \cite{BergNeuhaus91,MarinariParisi92,WangLandau01,NeuhausMagieraHansmann07}
To further check the correctness of the phase diagram, we therefore implement a code based on the
Wang-Landau algorithm \cite{WangLandau01} which is particularly suited to address first-order transitions.
This method performs a random walk in energy space and provides direct access to the density of states $\rho(E)$
(here, $E=L^2 e$ refers to the extensive total energy).
Specifically, the algorithm proceeds as follows:
We initially set $\rho(E)=1$. The energy histogram,
which records the visit to each energy level $E$, is started at $H(E)=0$. We then flip spins according to the probability
$p(E_i\to E_j)=\min(1,\rho(E_i)/\rho(E_j))$ where $E_i$ and $E_j$ are the energies of the states before and after the flip, respectively.
After every attempted spin flip, the density of states at the resulting energy is updated via $\rho(E) \to f \times \rho(E)$, and we
record the visit by updating the histogram, $H(E)\to H(E)+1$. The modification factor $f$ is initially set to $f=\exp(1)$.
Once the histogram $H(E)$ becomes reasonably flat, we reset $H(E)=0$ and update the modification factor to a smaller
value, $f\to f^{1/2}$. Iterating this procedure until $f<\exp(10^{-8})$  gives the density of
states $\rho(E)$ with high precision.
The right panel of Fig.~\ref{fig:P(m)_clean} shows the resulting density of states, weighted with the Boltzmann factor, for the clean Ashkin-Teller model
at $\epsilon=1$. The double peak structure characteristic of two coexisting phases is clearly visible.  The phase boundary can also be
estimated from the peak of the specific heat curve.

\begin{figure}
\includegraphics[width=8.7cm]{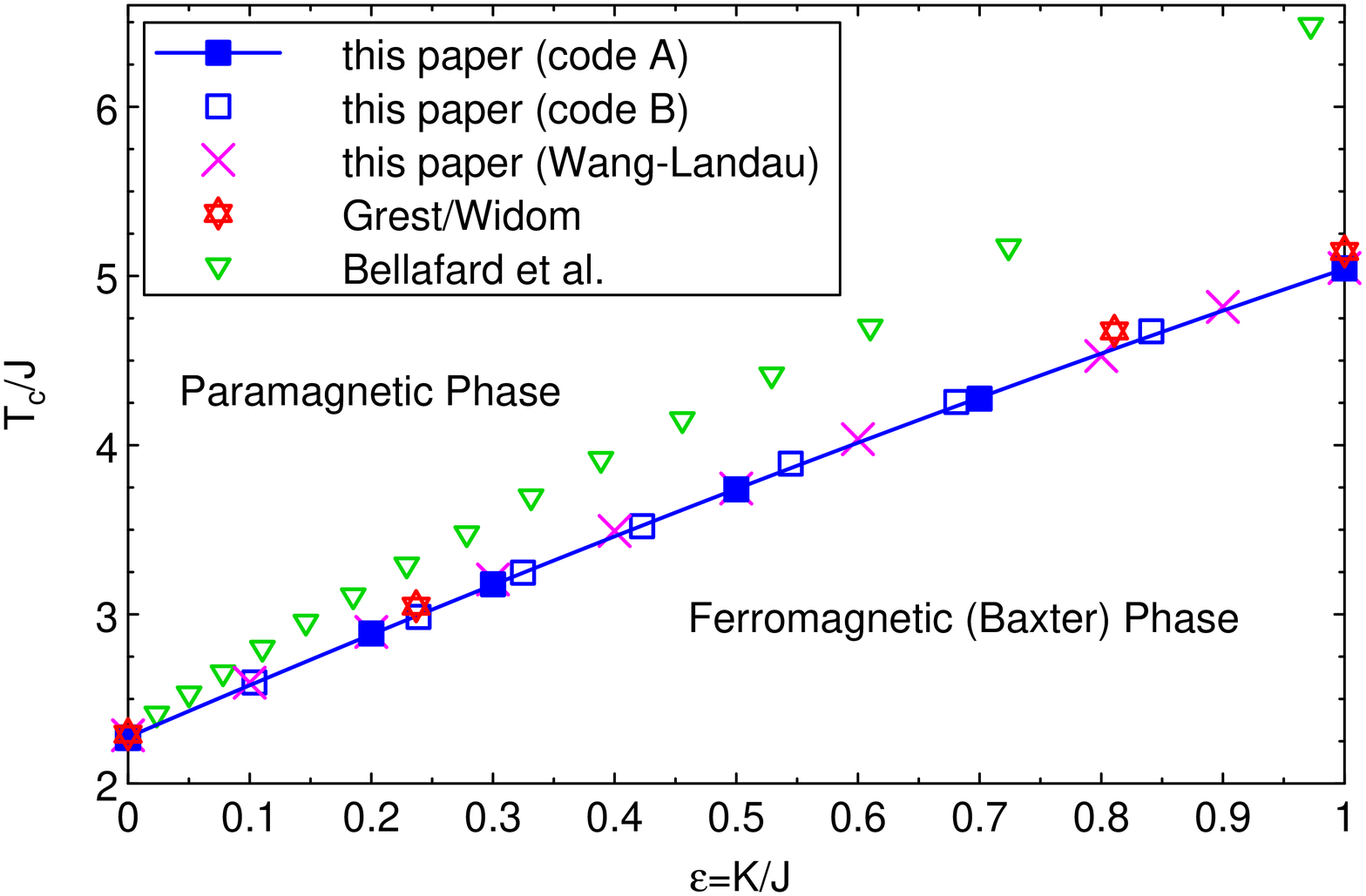}
\caption{(Color online) Phase diagram of the clean three-color Ashkin-Teller model as function of temperature $T$
         and coupling strength $\epsilon$. The blue squares and the pink crosses mark the numerically determined transition points while the line
         is just a guide to the eye. The error bars of all our data are significantly smaller than the symbol size.
         For comparison, the figure also shows data extracted from the papers by Grest and Widom \cite{GrestWidom81}
         and Bellafard et al.\cite{BellafardKatzgraberTroyerChakravarty12} }
\label{fig:pd_clean}
\end{figure}
The phase diagram presented in Fig.\ \ref{fig:pd_clean} summarizes the results of our calculations for the clean Ashkin-Teller model.
The figure also shows the critical temperatures reported in Refs.\ \onlinecite{GrestWidom81,BellafardKatzgraberTroyerChakravarty12}
(extracted by redigitizing Fig.\ 3 of Ref.\ \onlinecite{GrestWidom81} and Fig.\ 1 of Ref.\ \onlinecite{BellafardKatzgraberTroyerChakravarty12}).
Within the errors of the redigitized data, our results agree well with Ref.\ \onlinecite{GrestWidom81} but disagree with
Ref.\ \onlinecite{BellafardKatzgraberTroyerChakravarty12}.
\footnote{Our reproduction of the data of Ref.\ \onlinecite{BellafardKatzgraberTroyerChakravarty12} in Fig.\ \ref{fig:pd_clean} assumes
that the color sum in the Hamiltonian [eq.\ (1) of Ref.\ \onlinecite{BellafardKatzgraberTroyerChakravarty12}]
does not involve double counting of pairs of colors. If it does,
their $\epsilon$ values double, putting their phase boundary significantly to the right of ours.}

%%%%%%%%%%%%%%%%%%%%%%%%%%%%%%%%%%%%%%%%%%%%%%%%%%%%%%%%%%%%%%%%%%%%%%%%%%%%%%%%%
\subsection{Site-diluted Ising model}
\label{subsec:Results_Ising}
%%%%%%%%%%%%%%%%%%%%%%%%%%%%%%%%%%%%%%%%%%%%%%%%%%%%%%%%%%%%%%%%%%%%%%%%%%%%%%%%%

After discussing the clean limit $p=0, \epsilon \ne 0$, we now turn to the opposite limit
$p \ne 0, \epsilon=0$. In this limit, our Hamiltonian is equivalent to
three decoupled site-diluted Ising models.
The critical behavior of the disordered two-dimensional Ising model is actually an
interesting topic in itself because the clean correlation length exponent takes the
value $\nu=1$ which makes it marginal with respect to the Harris criterion \cite{Harris74}
$d\nu > 2$. In the literature, two main scenarios for the critical behavior have been put forward,
the logarithmic correction scenario and the weak-universality
scenario.

The logarithmic correction (strong-universality) scenario arises from  a
perturbative renormalization-group approach. \cite{DotsenkoDotsenko83,Shalaev84,Shankar87,*Shankar88,Ludwig88}
It predicts that the
asymptotic critical behavior of the disordered Ising model is controlled by the clean
Ising fixed point. Disorder, which is a marginally irrelevant operator,
gives rise to universal logarithmic corrections to scaling.
Specifically, one can derive the following finite-size scaling behavior \cite{MazzeoKuhn99,HTPV08,KennaRuizLorenzo08}
in the limit of large $L$.
The specific heat at the critical temperature diverges as
\begin{equation}
C \sim \ln \ln L~
\label{eq:C_lnln}
\end{equation}
with system size.
Magnetization and magnetic susceptibility at $T_c$ behave as
\begin{eqnarray}
M  &\sim& L^{-\beta/\nu}\, [1+O(1/(\ln L))]~,
\label{eq:M_ln}\\
\chi &\sim& L^{\gamma/\nu}\,[1+O(1/(\ln L))]~,
\label{eq:chi_ln}
\end{eqnarray}
with $\gamma/\nu=7/4$ and $\beta/\nu=1/8$ as in the clean Ising model. Any quantity $R$ of scale dimension zero
(such as the Binder cumulants $g_\textrm{av}$ and $g_\textrm{gl}$
as well as the correlation length ratios $\xi_\textrm{av}/L$ and $\xi_\textrm{gl}/L$)
and its temperature derivative scale as
\begin{eqnarray}
R &=& R^\ast + O(1/(\ln L))~,
\label{eq:R_ln}\\
dR/dT  &\sim& L^{1/\nu}(\ln L)^{-1/2}\, [1+O(1/(\ln L))]
\label{eq:dRdT_ln}
\end{eqnarray}
with $\nu=1$. This means, $\chi$, $M$, and $R$ do not have multiplicative logarithmic corrections
but $dR/dT$ has a multiplicative $(\ln L)^{-1/2}$ correction.

The weak-universality scenario was developed heuristically based on early numerical data. \cite{FahnleHoleyEckert92,KimPatrascioiu94,Kuhn94}
It states that the observables display simple power-law critical singularities. Their exponents
vary continuously with disorder strength, but certain ratios stay constant at their clean values,
for example $\gamma/\nu$ and $\beta/\nu$.
The debate over the critical behavior of the two-dimensional disordered Ising model has persisted
over many years, mainly because it is very hard to discriminate between logarithms and small powers on the
basis of numerical data. Only in the last few years, the evidence seems to favor
the logarithmic correction scenario (see, e.g., Refs.\
\onlinecite{HTPV08,KennaRuizLorenzo08,FytasMalakisHadjiagapiou08,FytasMalakis10}
and references therein).

The purpose of our simulations is twofold: On the one hand, the disordered Ising model is
an important (limiting) reference case for our main topic, the disordered Ashkin-Teller model.
On the other hand, we hope to make a contribution towards resolving the above controversy about
the disordered Ising model itself. We therefore perform a series of high-accuracy simulations for dilution
$p=0.3$ and $\epsilon=0$, using linear system sizes from $L=50$ to 2240.  The numbers of disorder
realizations range from $4 \times 10^6$ for $L=50$ to $5 \times 10^5$ for $L=2240$).

\begin{figure}
\includegraphics[width=8.7cm]{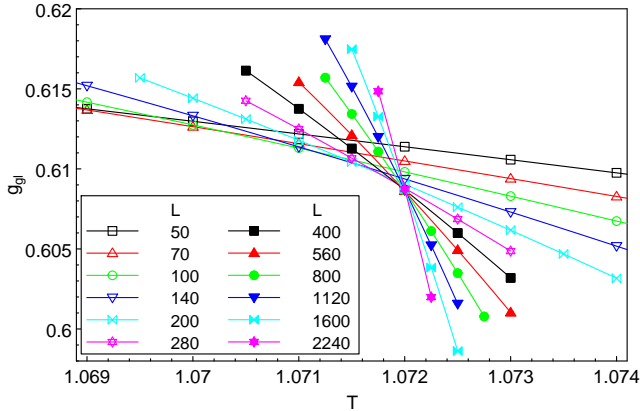}
\caption{(Color online) Binder cumulant $g_\textrm{gl}$ vs.\ temperature $T$ for $p=0.3$ and $\epsilon=0$
        for different linear system sizes $L$. The statistical errors are much smaller than the symbol size.
        With increasing $L$, the crossing point shifts towards higher $T$,
        indicating significant corrections to scaling.}
\label{fig:binderp03epsj00}
\end{figure}

\begin{figure}
\includegraphics[width=8.7cm]{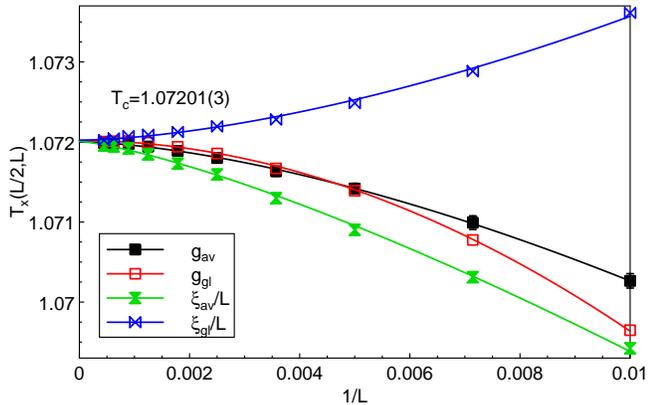}
\caption{(Color online) Crossing temperatures $T_x$ vs.\ inverse system size $1/L$ for $p=0.3$ and $\epsilon=0$.
$T_x$ is the temperature where the curves of $g_\textrm{av}, g_\textrm{gl},
\xi_\textrm{av}/L$ and $\xi_\textrm{gl}/L$ versus $T$ cross
for system sizes $L/2$ and $L$. The solid lines are fits to $T_x = T_c + a L^{-b}$.
The error bars of $T_x$ are about the size of the symbols at the right side of the plot
and become much smaller towards the left.}
\label{fig:crossingsp03epsj00}
\end{figure}

Figure \ref{fig:binderp03epsj00} shows the Binder cumulant $g_\textrm{gl}$ as a function of
temperature.
Analogous plots can be produced for $g_\textrm{av}$, $\xi_\textrm{gl}/L$ and $\xi_\textrm{av}/L$.
All these quantities display significant corrections to scaling manifest in the shift of
the crossing temperature with increasing $L$. To extrapolate to infinite system size, we determine
the crossings of the $g_\textrm{gl}$ vs.\ $T$ curves for sizes $L/2$ and $L$ (and the corresponding
crossings for $g_\textrm{av}$, $\xi_\textrm{gl}/L$ and $\xi_\textrm{av}/L$).
Figure \ref{fig:crossingsp03epsj00} presents the dependence of the crossing temperatures $T_x$ on
the system size.
The critical temperature $T_c$ can be extracted by extrapolating the crossing temperatures
to infinite system size. Fits to $T_x = T_c + a L^{-b}$ yield $T_c=1.07201(3)$ which agrees reasonably well with the result of
Ref.\ \onlinecite{HTPV08}, $T_c=1.07194(6)$, obtained from systems with up to $256^2$ sites. (Fits of $T_x$ vs.\ $1/L$ to quadratic polynomials
give comparable results.)

\subsubsection{Logarithmic-correction scenario}

To study the critical behavior, we analyze the system-size dependence  at the critical temperature
of magnetization $M$, susceptibility $\chi$, specific heat $C$
and the slope $d\ln(\xi_\textrm{gl}/L)/dT$ of the
normalized correlation length curves.
Straight power-law fits (without corrections to scaling) of the data at $T=1.07200$ to
$M \sim L^{-\beta/\nu}$, $\chi \sim L^{\gamma/\nu}$, $C \sim L^{\alpha/\nu}$ and
$d\ln(\xi_\textrm{gl}/L)/dT \sim L^{1/\nu}$
give the estimates $\beta/\nu=0.1217(1)$, $\gamma/\nu=1.8046(2)$, $\alpha/\nu=0.0516(1)$,
and $\nu= 1.107(3)$. These values do not agree with the clean Ising exponents,
$\beta/\nu=1/8$, $\gamma/\nu=7/4$, $\alpha/\nu=0$,
and $\nu= 1$.
However, the quality of all fits is extremely poor,
with reduced $\bar \chi^2$ values of about 50, 1300, 4600, and 7, respectively. This indicates that the data significantly
deviate from pure power laws.

\begin{figure}
\includegraphics[width=8.7cm]{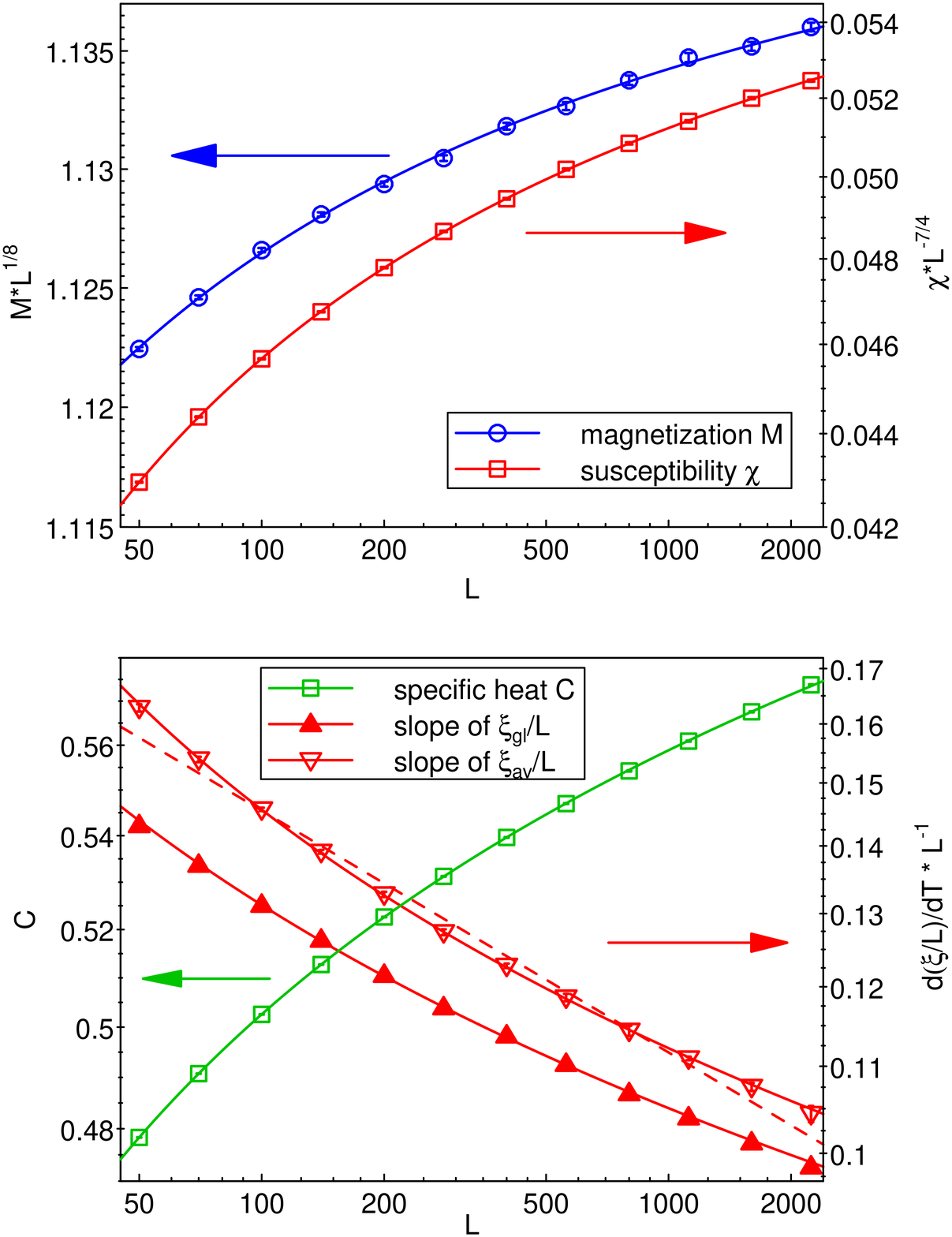}
\caption{(Color online) System-size dependence of observables at the critical temperature.
        Top: Log-log plots of $M L^{1/8}$ and $\chi L^{-7/4}$ vs.\ $L$ at $T=1.07200$
        for $p=0.3$ and $\epsilon=0$. The solid lines are fits to $a[1+b/\ln(cL)]$ as suggested by eqs.\
        (\ref{eq:M_ln}) and (\ref{eq:chi_ln}).
        Bottom: Log-log plots of the specific heat $C$ as well as the slopes $L^{-1}d\ln(\xi_\textrm{av}/L)/dT$ and $L^{-1}d\ln(\xi_\textrm{gl}/L)/dT$
        vs.\ $L$ at $T=1.07200$ for $p=0.3$ and $\epsilon=0$. The solid lines represent
        fits to  $a \ln[b \ln(cL)]$ for the specific heat and to $a [\ln(bL)]^{-1/2}$ for the slopes. The dashed line
        shows a power-law fit using $\nu=1.130$ as implied by the hyperscaling relation $2-\alpha = 2\nu$ in the
        weak-universality scenario.}
\label{fig:criticalityp03epsj00}
\end{figure}

To understand the nature of the deviations, we divide out the
clean Ising power laws and plot the resulting data in
Fig.\ \ref{fig:criticalityp03epsj00}.
$M L^{1/8}$ and $\chi L^{-7/4}$, shown in the upper panel, clearly increase much more slowly than power laws with $L$. As suggested in eqs.\ (\ref{eq:M_ln}) and (\ref{eq:chi_ln}),
their behaviors can be analyzed as logarithmic corrections to scaling and fitted to the form $a[1+b/\ln(cL)]$. The fits are of good quality (reduced $\bar \chi^2$ of 1.1 and 0.5 for the
magnetization and susceptibility, respectively).
The lower panel of Fig.\ \ref{fig:criticalityp03epsj00} shows specific heat $C$ vs.\ system size at the critical temperature.
The data can be fitted well by the double-logarithmic form $a \ln[b \ln(cL)]$ suggested by eq.\ (\ref{eq:C_lnln}), giving a reduced $\bar \chi^2$ of about 1.2.
For comparison, we also consider a simple logarithmic form $C=a\ln(bL)$. A semi-log plot of $C$ vs.\ $\ln(L)$ (not shown) shows strong deviations from a
straight line. Correspondingly, the simple logarithmic fit is  of very poor quality, with a reduced $\bar \chi^2$ of about 3000;
and it does not improve much if the fit range is restricted.

The lower panel of Fig.\ \ref{fig:criticalityp03epsj00} also shows the slopes $d\ln(\xi_\textrm{av}/L)/dT$ and
$d\ln(\xi_\textrm{gl}/L)/dT$ of the normalized correlation
length curves at the critical temperature. We again divide out the clean Ising power law $d\ln(\xi/L)/dT \sim L$ to make the
deviations from power-law behavior clearly visible. The resulting data can be fitted well by the logarithmic form
$a [\ln(bL)]^{-1/2}$ suggested by eq.\ (\ref{eq:dRdT_ln}) (reduced $\bar \chi^2$ of about 0.3 for both data sets).
Including extra additive corrections to scaling does not improve the fits.
The slopes of the magnetic Binder cumulants behave analogously.

\begin{figure}
\includegraphics[width=8.7cm]{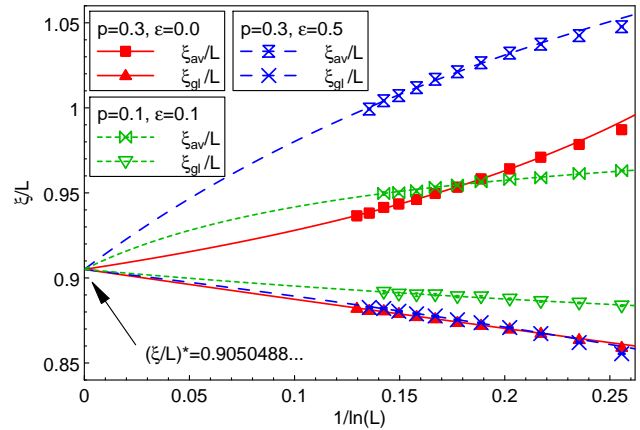}
\caption{(Color online) Dimensionless ratios  $\xi_\textrm{gl}/L$ and $\xi_\textrm{av}/L$ at criticality
        vs.\ $1/\ln(L)$ for $p=0.3,\epsilon=0$ ($T=1.07200$), $p=0.3,\epsilon=0.5$ ($T=1.93471$), and for
        $p=0.1,\epsilon=0.1$ ($T=2.18769$). The error bars are significantly smaller
        than the symbol size. The lines are fits
        to   $(\xi/L) = (\xi/L)^\ast + a/\ln(bL)$ with $(\xi/L)^\ast$ fixed at the clean Ising value
        0.9050488292. Note that the two $\xi_\textrm{gl}/L$ curves for $p=0.3,\epsilon=0$ and $p=0.3,\epsilon=0.5$
        are almost on top of each other.}
\label{fig:xiratiosp03epsj00}
\end{figure}

In addition to the quantities shown in Fig.\ \ref{fig:criticalityp03epsj00},  we study
the system-size dependence of the dimensionless ratio $\xi/L$ at criticality.
Within the logarithmic correction scenario, this ratio is expected to approach the universal
value $(\xi/L)^\ast$ of the clean Ising model which is known with high precision
(see, e.g., Ref.\ \onlinecite{SalasSokal00}). For a square lattice with periodic boundary conditions
(torus topology), it reads $(\xi/L)^\ast = 0.9050488292(4)$. According to eq.\ (\ref{eq:R_ln}),
the approach to this value is logarithmically slow. We therefore plot $\xi_\textrm{gl}/L$ and $\xi_\textrm{av}/L$
at the critical temperature as functions of $1/\ln(L)$ in Fig.\ \ref{fig:xiratiosp03epsj00}.
Both ratios can be well fitted to the form $(\xi/L) = (\xi/L)^\ast + a/\ln(bL)$, as suggested
by eq.\ (\ref{eq:R_ln}), with $(\xi/L)^\ast$ fixed
at the clean Ising value. The fits are of excellent quality (reduced $\bar \chi^2$ of 0.7 and 0.6, respectively)
if the fit range is restricted to system sizes $L>70$. We attribute the small deviations for the smallest $L$
to subleading terms\cite{HTPV08}  of the form $\ln\ln(bL)/\ln^2(bL)$ in eq.\ (\ref{eq:R_ln}) that are not included in the fit.
Analyzing the system size dependence of the Binder cumulant $g_\textrm{av}$ at criticality gives the same result:
$g_\textrm{av}$ approaches the universal clean Ising value \cite{SalasSokal00}, $g^\ast=0.610692(2)$
following eq.\ (\ref{eq:R_ln}).  The behavior of  $g_\textrm{gl}$ is more complex. With increasing $L$, it first
decreases below $g^\ast$ before turning around and approaching $g^\ast$ from below.
A quantitative analysis therefore requires even larger systems than ours to properly fit the subleading terms
in eq.\ (\ref{eq:R_ln}).

Finally, we also study the product order parameter
$M_p$. As the different colors are completely independent for $\epsilon=0$, $M_p$ must scale as $M^2$.
Indeed, the system size dependence of our data at criticality (not shown) can be fitted very well by $M_p = a L^{-1/4} [1+b/\ln(cL)]$.

In sum, our high-accuracy data almost perfectly agree with the renormalization-group predictions that lead to the
logarithmic correction scenario outlined in eqs.\ (\ref{eq:C_lnln}) to (\ref{eq:dRdT_ln}) over the entire range of system
sizes studied ($L=50$ to 2240).

\subsubsection{Weak-universality scenario}

Can these data also be understood within the heuristic weak-universality scenario? Figure \ref{fig:criticalityp03epsj00}
shows that the data deviate significantly from pure power laws over entire system size range.
The weak-universality scenario can thus only work, if at all, if corrections to scaling are included (in addition
to potential changes in the critical exponents).

The system size dependencies of $M$ and $\chi$ at $T_c$ shown
in Fig.\ \ref{fig:criticalityp03epsj00} can be fitted with the clean Ising exponents $\beta/\nu=1/8$ and
$\gamma/\nu=1$, provided that corrections to scaling of the type $M = a L^{-\beta/\nu}(1+b L^{-\omega})$ and
$\chi=a L^{\gamma/\nu}(1+b L^{-\omega})$ are included. These fits are of lower, but still acceptable, quality (reduced $\bar \chi^2$
of about 1.9 and 2.5, respectively) than the fits with
logarithmic corrections given in eqs.\ (\ref{eq:M_ln}) and (\ref{eq:chi_ln}).
Four-parameter fits to the same functional forms but with floating critical exponents give $\beta/\nu=0.123(10)$
and $\gamma/\nu=1.76(2)$ where the errors mostly stem from the sensitivity of the fits towards changes of the fit interval.
We conclude that $\beta/\nu$ and $\gamma/\nu$ agree with the clean Ising values. As these exponent ratios
are expected to take the clean values in both scenarios, this does not allow us to discriminate between the scenarios.
We therefore turn to the exponents $\alpha/\nu$ and $\nu$
which are expected to be nonuniversal.

In the weak-universality scenario, the
slow increase of the specific heat $C$ with $L$ (as shown in the lower panel of Fig.\ \ref{fig:criticalityp03epsj00})
is interpreted as power-law behavior of the type $C = C_\infty + a L^{\alpha/\nu}$ with a negative exponent $\alpha$.
%This means that $C$ takes the finite value $C_\infty$ at criticality.
A fit to this form yields $\alpha/\nu=-0.230(3)$, however, it is of much lower quality (reduced $\bar \chi^2$ of about 26)
than the double-logarithmic fit employed above. Moreover, the fit is very unstable. If we extract an effective exponent
by restricting the fit to the interval $(L_\textrm{min},4L_\textrm{min})$, its value increases monotonically from
$-0.278$ for $L_\textrm{min}=50$ to $-0.106$ for $L_\textrm{min}=560$, with no sign of saturation. A controlled extrapolation to
infinite $L_\textrm{min}$ is difficult; but it appears to be compatible with $\alpha/\nu=0$. In fact, a (perhaps overambitious) five-parameter
fit that includes subleading corrections to scaling, $C = C_\infty + a L^{\alpha/\nu} (1+bL^{-\omega})$, yields
a very small $\alpha/\nu \approx 0.03$ albeit with a large error of about 0.2.

The temperature derivatives $d\ln(\xi_\textrm{av}/L)/dT$ and $d\ln(\xi_\textrm{gl}/L)/dT$ of the normalized correlation
length curves at the critical temperature can be fitted with the clean Ising exponent $\nu=1$ if corrections to scaling
of the type $a L^{1/\nu}(1+b L^{-\omega})$ are included. The quality of these fits (reduced $\bar \chi^2$ of 0.5 and 0.2) is
comparable to that of the logarithmic fits above (note, however, that the logarithmic fits contained only two free parameters).
Four-parameter fits to the same functional form but with floating $\nu$ give $\nu=1.07(4)$ and 1.06(6) for the
average and global correlation length data, respectively.

If we ignore the strong size-dependence of the effective specific heat exponent and take
the value $\alpha/\nu=-0.230(3)$ resulting from the global fit, the hyperscaling relation $2-\alpha=2\nu$ yields a
correlation length exponent of $\nu=1.130(3)$. As the lower panel of Fig.\ \ref{fig:criticalityp03epsj00} shows, the
$d\ln(\xi/L)/dT$ data are clearly incompatible with this value, even if corrections to scaling are included.
\footnote{Interestingly, the correlation length exponent resulting from the hyperscaling relation, $\nu=1.130(3)$,
is not too far from the result of a naive power law fit of the  $d\ln(\xi_\textrm{gl}/L)/dT$ data which gives
$\nu=1.107(3)$. This approximate agreement of the effective exponents may explain why literature data on smaller systems
could be successfully fitted with the weak-universality scenario.}

Finally, we point out that the subleading exponent $\omega$ appearing in all the power-law fits is not very robust.
Its values seem to cluster around 0.35 but they vary between about 0.2 and 0.7 upon changing the fit intervals and
between quantities.

We conclude that the weak-universality scenario is not compatible with our numerical results:
Simple power-law singularities do not describe the data at all. If corrections to scaling are included,
we do not find evidence for the asymptotic exponents to be
different from the clean Ising ones.

\subsubsection{Other dilutions}

We have performed analogous simulations for dilution $p=0.2$, using systems with $50^2$
to $1120^2$ sites. The data are averaged over $10^5$ to $10^6$ disorder configurations.
By extrapolation the crossing temperatures of the Binder cumulant and the normalized correlation
length as above, we find a critical temperature of $T_c=1.50709(5)$. The system-size dependence
of observables at $T_c$ looks almost identical to that shown in Fig.\ \ref{fig:criticalityp03epsj00}
for $p=0.3$: The data feature pronounced deviations from power-law behavior that can be fitted
very well by the logarithmic-correction scenario, eqs.\ (\ref{eq:C_lnln}) to (\ref{eq:dRdT_ln}), over
the entire range of system sizes studied (the reduced $\bar \chi^2$ range between 0.4 and 1.1).

%%%%%%%%%%%%%%%%%%%%%%%%%%%%%%%%%%%%%%%%%%%%%%%%%%%%%%%%%%%%%%%%%%%%%%%%%%%%%%%%%
\subsection{Site-diluted Ashkin-Teller model}
\label{subsec:Results_diluted_AT}
%%%%%%%%%%%%%%%%%%%%%%%%%%%%%%%%%%%%%%%%%%%%%%%%%%%%%%%%%%%%%%%%%%%%%%%%%%%%%%%%%

%\subsubsection{Transition temperature and order of the transition}

After having discussed the limiting cases, we now turn to the
full problem, the site-diluted Ashkin-Teller model. We first consider a system with dilution $p=0.3$ and
four-spin coupling $\epsilon=0.5$ because we expect deviations from the clean Ising critical behavior,
if any, to be more easily visible if $p$ and $\epsilon$ are large.

\begin{figure}
\includegraphics[width=8.7cm]{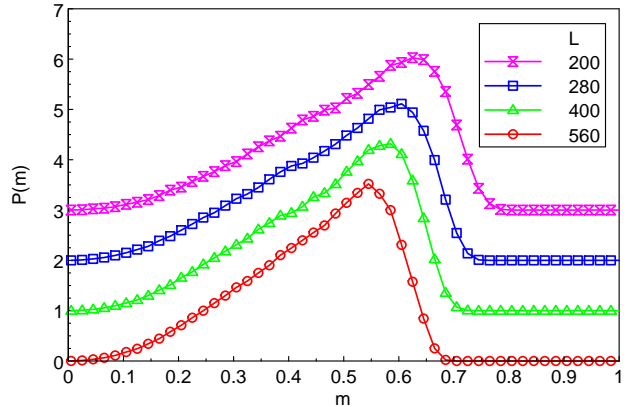}
\caption{(Color online) Magnetization distribution $P(m)$ of the three-color Ashkin-Teller model
with dilution $p=0.3$ and coupling $\epsilon=0.5$ close to the transition temperature $T_c \approx 1.94$. The
distribution features a single peak characteristic of a continuous transition.
(The curves for $L=400,280$, and 200 are shifted upwards by multiples of 1.0 for clarity.) }
\label{fig:P(m)_p03epsj05}
\end{figure}

According to the Aizenman-Wehr theorem,\cite{AizenmanWehr89} the first-order phase transition of the
clean system should be destroyed by dilution. We confirm this by calculating the magnetization distribution
$P(m)$ close to the transition temperature for systems of $200^2$ to $560^2$ sites (1000 disorder configurations
each). It is shown in Fig.\ \ref{fig:P(m)_p03epsj05}.
The distribution features a single broad peak characteristic of a continuous transition,
in contrast to the double-peak structure of the clean case (Fig.\ \ref{fig:P(m)_clean}).

\begin{figure}
\includegraphics[width=8.7cm]{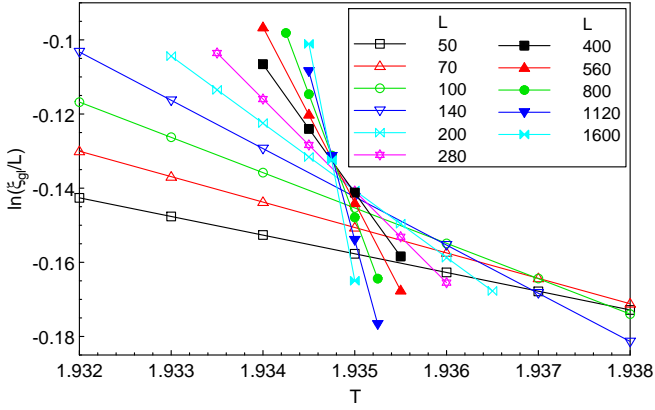}
\caption{(Color online) Normalized correlation length $\xi_\textrm{gl}/L$ vs.\ temperature $T$ for $p=0.3$ and $\epsilon=0.5$
        for different linear system sizes $L$. The statistical errors are significantly smaller than the symbol size.
        With increasing $L$, the crossing point shifts towards lower $T$,
        indicating significant corrections to scaling.}
\label{fig:xip03epsj05}
\end{figure}

\begin{figure}
\includegraphics[width=8.7cm]{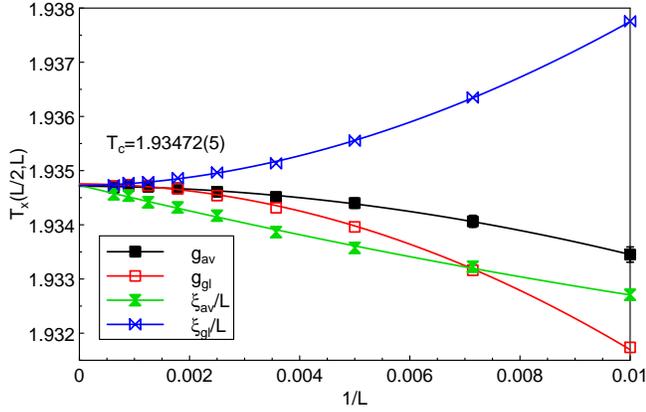}
\caption{(Color online) Crossing temperatures $T_x$ vs.\ inverse system size $1/L$ for $p=0.3$ and $\epsilon=0.5$.
$T_x$ is the temperature where the curves of $g_\textrm{av}, g_\textrm{gl},
\xi_\textrm{av}/L$ and $\xi_\textrm{gl}/L$ versus $T$ cross
for system sizes $L/2$ and $L$. The solid lines are fits to $T_x = T_c + a L^{-b}$.
The error bars of $T_x$ are about the size of the symbols at the right side of the plot
and become much smaller towards the left.}
\label{fig:crossingsp03epsj05}
\end{figure}

We then perform a series of high-accuracy simulations for linear system sizes from $L=50$ to 1600.
The numbers of disorder configurations range from $3\times 10^6$ for $L=50$ to $5\times 10^5$ for $L=1600$.
To find the critical point, we study the Binder cumulants $g_\textrm{av}(T)$ and $g_\textrm{gl}(T)$
as well as the normalized correlation lengths $\xi_\textrm{av}(T)/L$ and $\xi_\textrm{gl}(T)/L$.
The resulting data look qualitatively very similar to those of the diluted Ising model.
As an example, we present $\xi_\textrm{gl}/L$ in Fig.\ \ref{fig:xip03epsj05}.
The critical temperature can be estimated by extrapolating to infinite system size the temperatures where
the $g_\textrm{av}(T)$ curves (as well as the $g_\textrm{gl}(T)$, $\xi_\textrm{av}(T)/L$ and $\xi_\textrm{gl}(T)/L$ curves)
for sizes $L/2$  and $L$ cross.
Figure \ref{fig:crossingsp03epsj05} shows the system-size dependence of the crossing temperatures $T_x$.
Fits to $T_x = T_c + a L^{-b}$ yield the estimate $T_c=1.93472(5)$.
(Fits of $T_x$ vs.\ $1/L$ to quadratic polynomials give comparable results.)

\subsubsection{Critical behavior: Ising with logarithmic corrections}
\label{subsubsec:AT_log}

To analyze the critical behavior, we now study the system-size dependence  at criticality of
magnetization $M$, susceptibility $\chi$, specific heat $C$, and the slope
$d\ln(\xi_\textrm{gl}/L)/dT$ of the normalized correlation length.
Simple power-law fits over the entire system size range ($L=50$ to 1600) of the data at $T=1.93471$ to
$M \sim L^{-\beta/\nu}$, $\chi \sim L^{\gamma/\nu}$, $C \sim L^{\alpha/\nu}$ and
$d\ln(\xi_\textrm{gl}/L)/dT \sim L^{1/\nu}$
give the estimates $\beta/\nu=0.1238(1)$, $\gamma/\nu=1.7948(3)$, $\alpha/\nu=0.0735(1)$,
and $\nu= 1.075(2)$. These values do not agree with the clean Ising exponents,
but the quality of the fits is again very poor.
The reduced $\bar \chi^2$ values are about 21, 180, 4300, and 7, respectively, indicating systematic
deviations from pure power-law behavior.

\begin{figure}
\includegraphics[width=8.7cm]{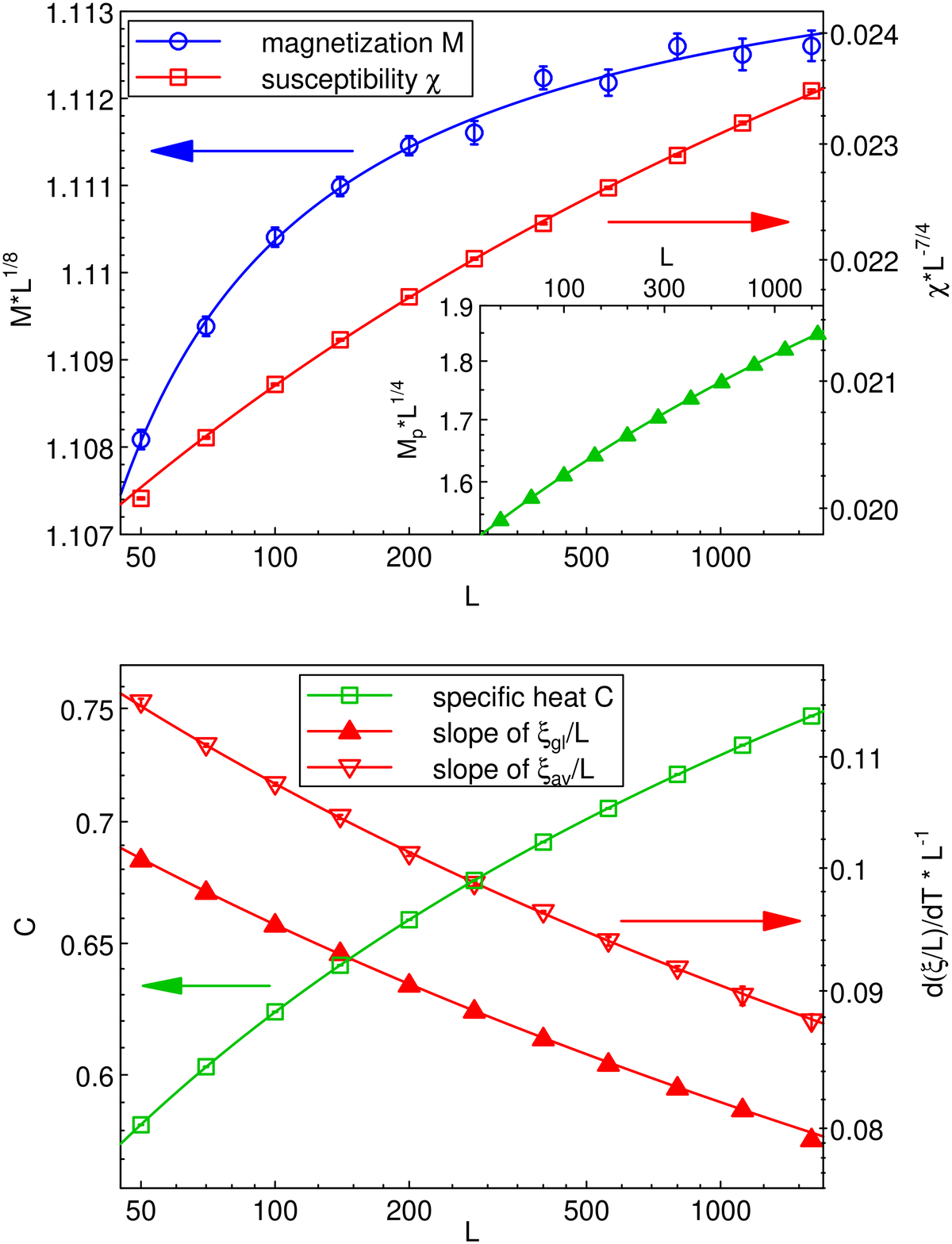}
\caption{(Color online) System-size dependence of observables at the critical temperature.
        Top: Log-log plots of $M L^{1/8}$ and $\chi L^{-7/4}$ vs.\ $L$ at
        $T=1.93471$ for $p=0.3$ and $\epsilon=0.5$. Inset: Log-log plot of $M_p L^{1/4}$ vs.\ $L$.
        All solid lines are fits to $a[1+b/\ln(cL)]$.
        Bottom: Log-log plots of the specific heat $C$ as well as the slopes $L^{-1}d\ln(\xi_\textrm{av}/L)/dT$ and $L^{-1}d\ln(\xi_\textrm{gl}/L)/dT$
        vs.\ $L$ at $T=1.93471$ for $p=0.3$ and $\epsilon=0$. The solid lines represent
        fits to  $a \ln[b \ln(cL)]$ for the specific heat and to $a [\ln(bL)]^{-1/2}$ for the slopes. }
\label{fig:criticalityp03epsj05}
\end{figure}

To investigate these deviations in detail, we proceed analogously to the diluted Ising model
in Sec.\ \ref{subsec:Results_Ising}, i.e., we divide out the clean Ising critical behavior
and present the resulting data in Fig.\ \ref{fig:criticalityp03epsj05}.
The figure shows that none of the plotted quantities follow simple power laws; instead they vary more slowly with
$L$ over the entire system size range.

Motivated by Cardy's renormalization group \cite{Cardy96,Cardy99}, we therefore attempt to fit the data with
the clean Ising exponents and logarithmic corrections analogous to those of the diluted Ising model, eqs.\
(\ref{eq:C_lnln}) to (\ref{eq:dRdT_ln}).
The magnetization can be fitted well to the form $M= a L^{-1/8} [1+b/\ln(cL)]$ over the entire
size range $L=50$ to 1600. The reduced $\bar \chi^2$ is about 0.9. Fitting the susceptibility to
$\chi= a L^{7/4} [1+b/\ln(cL)]$ over the entire size range leads to an unsatisfactory reduced $\bar \chi^2 \approx 7$.
However, the fit becomes of good quality (reduced $\bar \chi^2 \approx 1.8$) if we drop the two smallest system sizes,
restricting the fit to the range $L=100$ to 1600. We attribute this to the crossover from the strong first-order
transition in the clean case to our critical point. (This crossover will be studied in detail in Sec.\ \ref{subsec:crosover}.)

We also analyze the product order parameter $M_p$. Within Cardy's theory,\cite{Cardy96,Cardy99} the critical
renormalization group fixed point is at
$\epsilon=0$. This means different colors decouple at criticality. Thus, $M_p$ should scale as $M^2$.
In agreement with this expectation, the system size dependence of our data at criticality (shown in the inset)
can be fitted by $M_p = a L^{-1/4} [1+b/\ln(cL)]$, giving a reduced $\bar \chi^2$ of about 1.

The specific heat, shown in the lower panel of Fig.\ \ref{fig:criticalityp03epsj05}, can be fitted well by the
double-logarithmic form $C=a\ln[b\ln(cL)]$ over the entire size range, giving a reduced $\bar \chi^2$ of a about 1.2.
Finally, the slopes $d\ln(\xi_\textrm{av}/L)/dT$ and $d\ln(\xi_\textrm{gl}/L)/dT$ of the normalized correlation lengths
at criticality can be fitted by the form $aL[\ln(bL)]^{-1/2}$ over the entire size range (reduced $\bar \chi^2$ of about
0.5 and 0.2, respectively).

Finally, we investigate the system size dependence of the dimensionless ratio $\xi/L$ at criticality.
Fig.\ \ref{fig:xiratiosp03epsj00} presents  $\xi_\textrm{gl}/L$ and $\xi_\textrm{av}/L$ as functions of $1/\ln(L)$.
Both ratios can be well fitted to the form $(\xi/L) = (\xi/L)^\ast + a/\ln(bL)$ with $(\xi/L)^\ast$ fixed
at the clean Ising value, as suggested by  eq.\ (\ref{eq:R_ln}). The fits are of good quality (reduced $\bar \chi^2$ of 1.0 and 1.6,
respectively) if the fit range is restricted to system sizes $L>100$. The deviations for the smaller $L$ likely
stem from the crossover between the clean first-order
transition and our critical point as well as from subleading terms of the form $\ln\ln(bL)/\ln^2(bL)$ in
eq.\ (\ref{eq:R_ln}) that are not included in the fit.

We conclude that all our data can be described nearly perfectly in terms of the clean Ising critical behavior
with logarithmic corrections to scaling, as predicted by Cardy's renormalization group. \cite{Cardy96,Cardy99}

\subsubsection{Power law behavior?}

Even though our analysis does not show any disagreements between the Monte Carlo data and renormalization group predictions,
we still test whether the data are compatible with nonuniversal power-law critical behavior as suggested in Refs.\
\onlinecite{BellafardKatzgraberTroyerChakravarty12,BellafardChakravartyTroyerKatzgraber14}.
Since the quantities shown in Fig.\ \ref{fig:criticalityp03epsj05} do not follow simple power laws,
it is clear that corrections to scaling need to be included in addition to possible deviations of the exponents
from the clean Ising values.

Magnetization, susceptibility and the slopes of the normalized correlation lengths can be fitted to
$M = a L^{-\beta/\nu}(1+b L^{-\omega})$, $\chi=a L^{\gamma/\nu}(1+b L^{-\omega})$, and
$d\ln(\xi/L)/dT = a L^{-1/\nu}(1+b L^{-\omega})$.
If the exponents $\beta/\nu$ and $\gamma/\nu$ and $\nu$ are fixed at the clean
Ising values 1/8, 7/4 and 1, respectively, the fits are of good quality, with reduced $\bar \chi^2$
just slightly higher than the logarithmic fits above.
(The system size range for the susceptibility fit needs to be restricted to $L \ge 100$
to achieve an acceptable quality.)
Four-parameter fits over the entire system size range to the same functional forms, but
with floating $\beta/\nu$, $\gamma/\nu$, and $\nu$ give the values
$\beta/\nu=0.125(1)$ and $\gamma/\nu=1.78(1)$ and $\nu=1.04(6)$ where the errors
mostly stem from the sensitivity of the fits towards removing points from the ends of the system size range.
Note that $\beta/\nu$ and $\gamma/\nu$ do not quite fulfill the hyperscaling relation
$2 \beta/\nu + \gamma/\nu =2$, suggesting that these values are not the true asymptotic exponents.

It is worth pointing out that the effective inverse correlation length exponent
$1/\nu_\textrm{eff}$, obtained by fitting the correlation length slopes
over a finite system size range, is always smaller than unity.
This can be seen from the downward slope of  $L^{-1} d\ln(\xi/L)/dT$ vs.\ $L$
in the lower panel of Fig.\  \ref{fig:criticalityp03epsj05}; it also
directly follows from (\ref{eq:dRdT_ln}). The effective correlation length
exponent thus fulfills $\nu_\textrm{eff} > 1$.
In contrast to Refs.\
\onlinecite{BellafardKatzgraberTroyerChakravarty12,BellafardChakravartyTroyerKatzgraber14}, we see no indications
of the inequality $d\nu \ge 2$ due to Chayes et al.\cite{CCFS86} being violated even by the effective exponent.

The most interesting quantity is the specific heat. A correlation length exponent $\nu \ge 1$ implies,
 via the hyperscaling relation $2-\alpha=2\nu$, that the specific heat exponent $\alpha \le 0$.
We therefore attempt to fit the data to the form $C=C_\infty + a L^{\alpha/\nu}$. A fit of all system sizes
yields $\alpha/\nu=-0.170(4)$, but the reduced $\bar \chi^2 \approx 12$ is unacceptably large.  Moreover, the fit is unstable.
If we extract an effective exponent $(\alpha/\nu)_\textrm{eff}$
by restricting the fit interval to $(L_\textrm{min},4L_\textrm{min})$, its value varies between
$-0.204$ and $-0.131$ for $L_\textrm{min}$ between 50 and 400. Extrapolating these values
to infinite system size does not give a definite answer. Depending on the mathematical model used for the
extrapolation, we find values between $-0.12$ and 0.

We again point out that the subleading exponent $\omega$ appearing in all the power-law fits is not very robust.
Its values vary between about 0.2 and 1.1 upon changing the fit intervals and between quantities.

We conclude that the description of our data in terms of power-law singularities does not work nearly
as well as the logarithmic correction scenario of Sec.\ \ref{subsubsec:AT_log}. Simple power laws do not describe the data.
If we insist on fitting the data to power laws with corrections to scaling included, there is no compelling evidence
for the true asymptotic exponents to differ from the clean Ising values.

\subsubsection{Universality and crossover between the clean and dirty phase transitions}
\label{subsec:crosover}

We perform analogous simulations for several additional values of dilution $p$ and
coupling strength $\epsilon$ in order to test whether the asymptotic critical behavior is universal.
In addition, we wish to explore the interesting crossover from the first-order transition of the
clean Ashkin-Teller model to the critical point of the disordered system. As discussed at the end of Sec.\
\ref{Model}, it should be particularly pronounced when the first-order transition of the clean system
is strong and the disorder is weak.

\begin{figure}
\includegraphics[width=8.7cm]{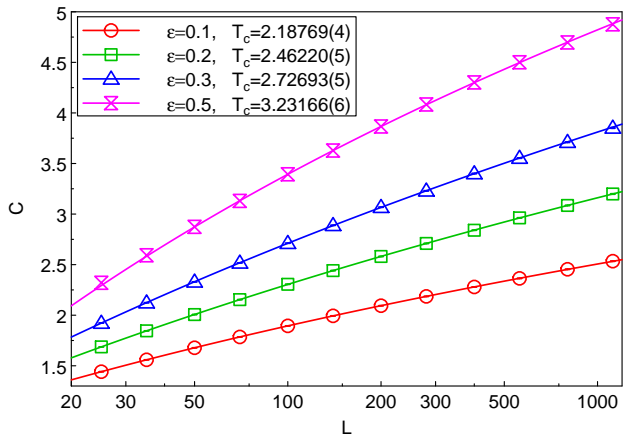}
\caption{(Color online) Semi-log plot of specific heat $C$ vs.\ system size $L$ at criticality
  for dilution $p=0.1$ and several couplings $\epsilon$. The error bars are much smaller than the symbol size.
  The solid lines are fits to $C=a \ln[b \ln(cL)]$.}
\label{fig:CV_pp01}
\end{figure}

To analyze the crossover, we therefore  perform a series of simulations for the weaker
dilution $p=0.1$. The coupling $\epsilon$ takes values 0.1, 0.2, 0.3, and
0.5. (Increasing $\epsilon$ increases the strength of the first-order transition in the
corresponding clean system.) These simulations use sizes between $L=25$ and 1120 with
$10^5$ to $10^6$ disorder realizations each. The data analysis follows the steps
outlined above, and the resulting critical temperatures are listed in the legend of Fig.\
\ref{fig:CV_pp01}.
Interestingly, the crossing temperature $T_x$ of the Binder cumulants and the correlation length ratios shifts much less with
system size than for $p=0.3$, indicating weaker disorder-induced corrections to scaling.
(For the crossings between the $L=50$ and $L=100$ curves, $(T_x-T_c)/T_c$ is roughly one order of magnitude smaller for $p=0.1$ than for $p=0.3$.)

Figure \ref{fig:CV_pp01} displays a semi-logarithmic plot of the specific heat $C$ at criticality vs.\ system size $L$.
For all $\epsilon$, $C$ curves downward, indicating that it increases more slowly than
logarithmic with $L$. The figure also shows fits to the double-logarithmic form C=$a \ln[b \ln(cL)]$ suggested by (\ref{eq:C_lnln}).
While the fits look nearly perfect to the eye, a $\bar \chi^2$ analysis reveals the
effects of the crossover from clean to dirty behavior: For $\epsilon=0.1$ and 0.2,
fits over the entire size-range $L=25$ to 1120 are of good quality (reduced $\bar \chi^2\approx 0.4$ and 0.7,
respectively). The quality decreases for $\epsilon=0.3$ (reduced $\bar \chi^2 \approx 2.3$) and
$\epsilon=0.5$ (reduced $\bar \chi^2 \approx 9$). Good quality fits (reduced $\bar \chi^2 < 2$) can be
restored by restricting the fit range to $L\ge 35$ for $\epsilon=0.3$ and to $L\ge 50$ for $\epsilon=0.5$.

\begin{figure}
\includegraphics[width=8.7cm]{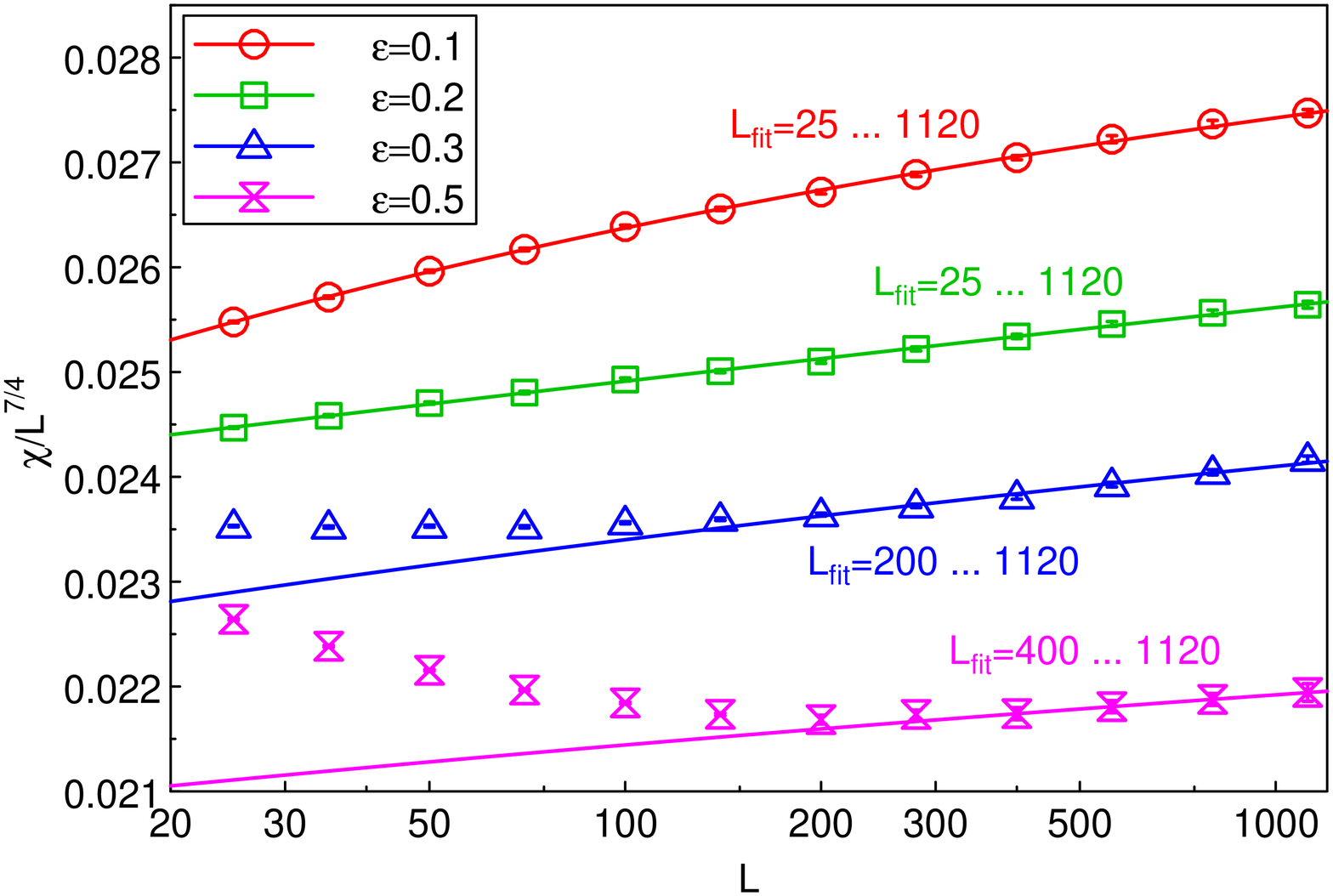}
\caption{(Color online) Semi-log plot of $\chi L^{-7/4}$ vs.\ $L$ at criticality
  for dilution $p=0.1$ and several couplings $\epsilon$. The data for $\epsilon=0.2$, 0.3 and 0.5
  are shifted upwards by 0.003, 0.005, and 0.008 for clarity. The error bars are much smaller than the symbol size.
  The solid lines are fits to $\chi/L^{7/4}= a[1+b/\ln(cL)]$, the fit ranges are indicated in the graph.}
\label{fig:chi_scaled_pp01}
\end{figure}

More pronounced signatures of the crossover from clean to dirty behavior can be found in the corrections
to the leading power-law size dependencies of magnetization and susceptibility (probably because these
corrections are weak -- they only change the observables by a few percent over the entire system-size
range). Figure \ref{fig:chi_scaled_pp01} presents $\chi/L^{7/4}$ vs.\ $L$ at criticality for dilution
$p=0.1$ and several $\epsilon$.
The data for $\epsilon=0.1$ behave analogously to those observed
earlier in Fig.\ \ref{fig:criticalityp03epsj05} for $p=0.3,\epsilon=0.5$. They can be fitted
well (reduced $\bar \chi^2 \approx 0.5$) with the logarithmic form  $\chi/L^{7/4}=a[1+b/\ln(cL)]$.
The same holds for the $\epsilon=0.2$ data which give a reduced $\bar \chi^2\approx 0.7$ (Note, however,
that the curvature of the $\epsilon=0.2$ data is very weak.)
For larger $\epsilon$, the behavior changes. $\chi/L^{7/4}$ first decreases with increasing $L$ before
turning around and starting to increase slowly. The increase can be fitted to
$a[1+b/\ln(cL)]$ for $L\ge 200$ ($\epsilon=0.3$) and $L\ge 400$ ($\epsilon=0.5$).

The corrections to the leading power-law behavior of the magnetization behave in the same fashion
as those of the susceptibility. We thus conclude that the asymptotic critical behavior is compatible
with the logarithmic correction scenario for all shown $\epsilon$ values.
We emphasize, however, that large system
sizes are necessary to reach
this asymptotic behavior if the first-order transition of the corresponding clean system
is strong ($\epsilon=0.3$ and 0.5) and the disorder is weak.
The universality of the critical behavior is also confirmed by the analysis of the ratio $\xi/L$
at criticality for $p=0.1,\epsilon=0.1$, shown in Fig.\ \ref{fig:xiratiosp03epsj00}.

In addition to the parameter sets already discussed, we also perform simulations for
$p=0.05, \epsilon=0.05$ as well as $p=0.3, \epsilon=0.3$ (system sizes between $L=50$ and 1120
with up to $10^6$ disorder realizations). In both cases, the critical behavior can be fitted well
with clean Ising critical behavior and logarithmic corrections to scaling over the entire system
size range.

%%%%%%%%%%%%%%%%%%%%%%%%%%%%%%%%%%%%%%%%%%%%%%%%%%%%%%%%%%%%%%%%%%%%%%%%%%%%%%%%%
\subsection{Random-bond Ashkin-Teller model}
\label{subsec:Results_randombond_AT}
%%%%%%%%%%%%%%%%%%%%%%%%%%%%%%%%%%%%%%%%%%%%%%%%%%%%%%%%%%%%%%%%%%%%%%%%%%%%%%%%%

The random-bond Ashkin-Teller model is expected to be in the same universality class as the site-diluted
model because both types of randomness are implementations of random-$T_c$ disorder. However, in view of the
unexpected results of Refs.\ \onlinecite{BellafardKatzgraberTroyerChakravarty12,BellafardChakravartyTroyerKatzgraber14},
we also perform a number of simulations for the random-bond case.

We employ code A to simulate systems using the binary bond distribution (\ref{eq:P(J)}) with
$J_h=2$, $J_l=0.5$ and concentration $c=0.5$. The four-spin interactions
$K_{ij}$ are slaved to the Ising interactions $J_{ij}$ via $K_{ij} = \epsilon J_{ij}$
with uniform $\epsilon$. We perform a series of runs for $\epsilon=0, 0.1, 0.2$, and $0.5$.
The linear system sizes are between $L=35$ and 1120; and the numbers of disorder realizations
for each parameter set range from $10^5$ for $L=1120$ to $10^6$ for $L=35$.

\begin{figure}
\includegraphics[width=8.7cm]{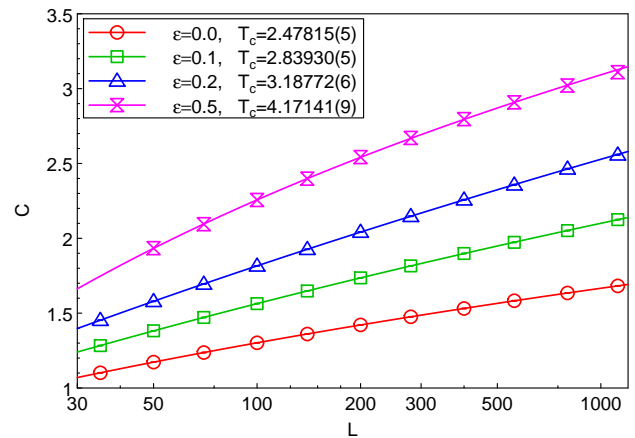}
\caption{(Color online) Semi-log plot of specific heat $C$ vs.\ system size $L$
  at criticality for the random bond-Ashkin-Teller model with $J_h=2, J_l=0.5$ and $c=0.5$.
  The error bars are much smaller than the symbol size.
  The solid lines are fits to $C=a \ln[b \ln(cL)]$.}
\label{fig:CV_rb05}
\end{figure}

The data analysis follows the steps outlined in Sec.\ \ref{subsec:Results_diluted_AT}.
The critical temperatures found by extrapolating the crossing temperatures
$T_x$ of the Binder cumulants $g_{av}$ and $g_{gl}$ as well as the correlation lengths ratios
$\xi_{av}/L$ and $\xi_{gl}/L$ to infinite system size are shown in the legend of Fig.\
\ref{fig:CV_rb05}.
We note that even though the bond randomness looks substantial ($J_h/J_l=4$), the disorder-induced
corrections to scaling turn out to be rather weak: The shifts of the crossing temperatures
$T_x$ with system size are even smaller than those for dilution $p=0.1$.
Because of the weaker corrections to scaling, effective exponents extracted by simple
power-law fits over the entire system size range are already very close to the expected
clean Ising values. For example, for $\epsilon=0.2$, we find
$\beta/\nu=0.1258(1), \gamma/\nu=1.7527(5)$, and $\nu=1.031(2)$.

We now analyze whether the asymptotic critical behavior can be described by logarithmic corrections
to the clean Ising power laws, as was the case for site dilution.
Figure \ref{fig:CV_rb05} displays a semi-logarithmic plot of the specific heat $C$ at criticality vs.\ system size $L$.
For all $\epsilon$, $C$ curves downward, indicating that it increases more slowly than
logarithmic with $L$. The figure also shows fits to the double-logarithmic form C=$a \ln[b \ln(cL)]$ suggested
by (\ref{eq:C_lnln}). All fits are of high quality (reduced $\bar \chi^2 <2$), for $\epsilon=0.5$ this requires
us to restrict the fit range to $L>100$.

\begin{figure}
\includegraphics[width=8.7cm]{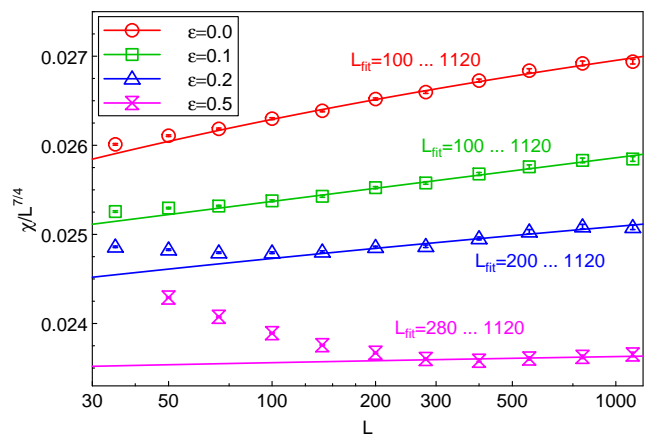}
\caption{(Color online) Semi-log plot of $\chi L^{-7/4}$ vs.\ $L$ at criticality
  for the random bond-Ashkin-Teller model with $J_h=2, J_l=0.5$ and $c=0.5$. The data for $\epsilon=0.1$, 0.2 and 0.5
  are shifted upwards by 0.004, 0.007, and 0.013 for clarity. The error bars are much smaller than the symbol size.
  The solid lines are fits to $\chi/L^{7/4}= a[1+b/\ln(cL)]$, the fit ranges are indicated in the graph.}
\label{fig:chi_scaled_rb05}
\end{figure}

The analysis of susceptibility and magnetization at criticality again reveal signatures of the crossover from clean
to dirty behavior.
 Figure \ref{fig:chi_scaled_rb05} presents $\chi/L^{7/4}$ vs.\ $L$ at criticality for different $\epsilon$.
The data for $\epsilon=0$ and 0.1 can be fitted to the logarithmic form
 $\chi/L^{7/4}=a[1+b/\ln(cL)]$ for sizes $L \ge 100$ (with reduced $\bar \chi^2 < 2$).
 For larger $\epsilon$,
  $\chi/L^{7/4}$ first shows a pronounced decrease with increasing $L$ before
turning around and starting to increase slowly. The increase can be fitted to
$a[1+b/\ln(cL)]$ for $L\ge 200$ ($\epsilon=0.2$) and $L\ge 280$ ($\epsilon=0.5$).
It must be noted however, that the susceptibility corrections to scaling in these data sets
are so weak (in agreement with the small shifts of $T_x$ mentioned above) that we
cannot unequivocally confirm their functional form. Their large-$L$ behavior
is certainly compatible with the predicted $a[1+b/\ln(cL)]$ form but other
functions would work as well. The corrections to the clean Ising power laws for the
magnetization behave analogously to those of the susceptibility.

\begin{figure}
\includegraphics[width=8.7cm]{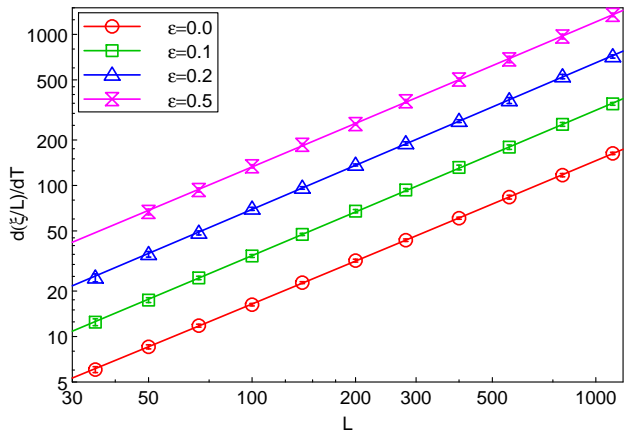}
\caption{(Color online) Log-log plot of the slopes $d\ln(\xi_\textrm{gl}/L)/dT$ vs. $L$ at criticality
  for the random bond-Ashkin-Teller model with $J_h=2, J_l=0.5$ and $c=0.5$. The data for $\epsilon=0.1$, 0.2 and 0.5
  are multiplied by factors 2, 4, and 8 for clarity. The error bars are much smaller than the symbol size.
  The solid lines are fits to $d\ln(\xi_\textrm{gl}/L)/dT = a L [\ln(bL)]^{-1/2}$.}
\label{fig:slopes_rb05}
\end{figure}

Finally, Fig.\ \ref{fig:slopes_rb05} shows the slopes $d\ln(\xi_\textrm{gl}/L)/dT$ of the correlation
length ratio vs.\ $L$ at criticality.
All curves can be fitted with high quality (reduced $\bar \chi^2 \lessapprox 1$) by the form
$a L  [\ln(bL)]^{-1/2}$ suggested by eq.\ (\ref{eq:dRdT_ln}). (Because the deviations from
the clean Ising power laws are again weak, the fits cannot 	unambiguously discriminate
between different functional forms: Simple power laws work as well, giving
exponents $\nu$ in the range of 1.03 to 1.05.) We conclude that the asymptotic critical behavior
of the random-bond Ashkin-Teller model is fully compatible with Cardy's predictions,
i.e., clean Ising exponents with logarithmic corrections.

In all of the above (code A) simulations, the four-spin interactions
$K_{ij}$ are slaved to the Ising interactions $J_{ij}$ via $K_{ij} = \epsilon J_{ij}$
with uniform $\epsilon$. In addition, we study random-bond Ashkin-Teller
models with uniform $K_{ij}\equiv K$ employing code B.
As summarized in Sec.\ \ref{subsec:MC_random_bond}, we consider $J_h=6/5$ and $J_l=4/5$ with equal probability $c=0.5$, implying an average interaction of
$J\equiv (J_h+J_l)/2 = 1$. Note that this disorder is much weaker than the disorder considered in the code-A simulations above.
The uniform, non-random four-spin interaction is given by $K=\epsilon J$ with $\epsilon=0.1$.
We simulate systems having linear sizes from $L=24$ to $1600$. The number of disorder realizations ranges from $10^4$ for $L=1600$ to $10^5$ for $L=24$.

\begin{figure}
\includegraphics[width=8.7cm]{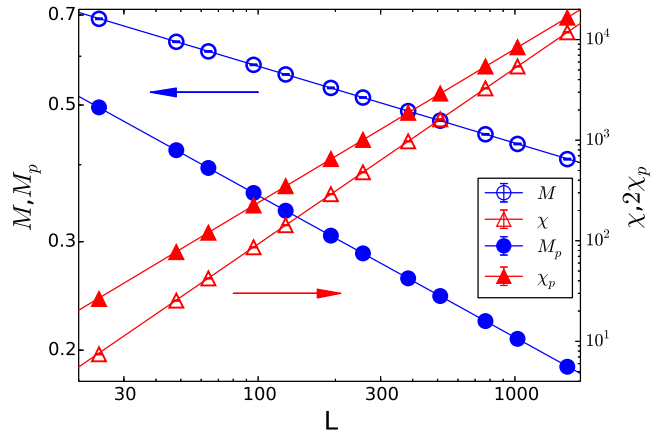}
\caption{(Color online) Log-log plot of the magnetization $M$, susceptibility $\chi$, polarization $M_p$ and polarization susceptibility $\chi_p$ vs.\ $L$ at $T_c$ for the random bond-Ashkin-Teller model with $J_h=6/5, J_l=4/5$ and uniform $K=0.1$. All error bars are much smaller than the symbol sizes. The solid lines are power-law fits.}
\label{fig:criticality_x01r15pow}
\end{figure}

The analysis of the data generated by code B proceeds as in Sec.\ \ref{subsec:Results_diluted_AT}.
The critical temperature is found by extrapolating the crossing temperature $T_x$ of the Binder cumulants $g_{av}$  to the
infinite system size limit. This yields a critical temperature $T_c=2.55625(1)$.
Simple power-law fits, shown in Fig.\ \ref{fig:criticality_x01r15pow}, of observables at $T_c$ to $M\sim L^{-\beta/\nu}$, $\chi \sim L^{\gamma/\nu}$, $M_p\sim L^{\beta_p/\nu}$ and $\chi_p \sim L^{\gamma_p/\nu}$
give $\beta/\nu=0.125(1)$, $\gamma/\nu=1.749(3)$, $\beta_p/\nu=0.230(1)$ and $\gamma_p/\nu=1.53(1)$.
% with reduced $\bar{\chi}^2$ 0.996, 0.723, 0.661 and 0.782.
The fits are of good quality (once again reduced $\bar \chi^2 <2$) if we restrict them to system sizes $L\geq 96$.
The exponents of magnetization and susceptibility have already locked onto the clean Ising values
$\beta/\nu=1/8$ and $\gamma/\nu=7/4$ within their error bars. The exponents related to $M_p$ and $\chi_p$ do not quite
agree with the expected values $\beta_p/\nu=2\beta/\nu=1/4$ and $\gamma_p/\nu=2-2\beta_p/\nu=3/2$, but they are close.
This is in tune with the results obtained for random-bond Ashkin-Teller model simulated via code A.

\begin{figure}
\includegraphics[width=8.7cm]{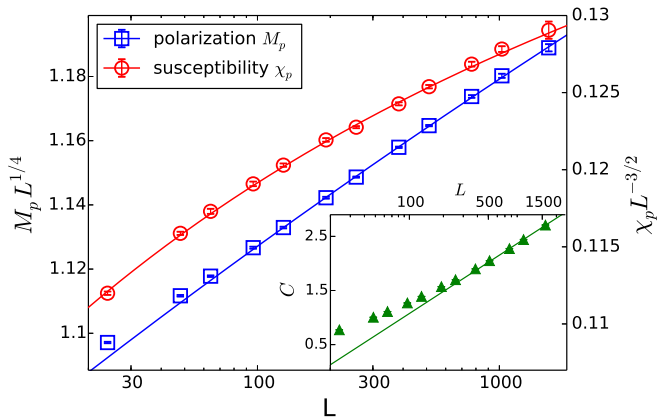}
\caption{(Color online) Semi-log plot of $M_p L^{1/4}$ and $\chi_p L^{-3/2}$ vs.\ $L$ at $T_c$ for the random bond-Ashkin-Teller model
with $J_h=6/5, J_l=4/5$ and uniform $K=0.1$.  The solid lines are fits to $a[1+b/\ln (cL)]$. Insert: Semi-log plot of the specific heat $C$
vs.\ $L$, the solid line is the fit to $a\ln(bL)$.}
\label{fig:criticality_x01r15log}
\end{figure}

Can the deviations of $M_p$ and $\chi_p$ from the expected behavior be explained by logarithmic corrections?
To answer this question, we again divide out the expected power laws and present the resulting data in
Fig.\ \ref{fig:criticality_x01r15log}.
The product order parameter, $M_p$ and the associated susceptibility $\chi_p$
can be fitted quite well with $M_p=aL^{-1/4}[1+b/\ln(cL)]$ and $\chi_p=aL^{3/2}[1+b/\ln(cL)]$.
For sizes $L \ge 96$, the reduced $\bar \chi^2$ are 1.26 and 0.713 for $M_p$, and $\chi_p$, respectively.

Corrections to the clean Ising behavior of magnetization and susceptibility are very weak
(in agreement with the fact that the exponents of simple power-law fits already coincide with the
clean Ising ones). If we include the smaller system sizes in the fits, these weak corrections cannot be fitted
satisfactorily with the universal logarithms (\ref{eq:M_ln}) and (\ref{eq:chi_ln}).
Similarly, the specific heat does not follow the double-logarithmic form, $C=a \ln[b \ln(cL)]$
(see inset of Fig.~\ref{fig:criticality_x01r15log}). Instead, the data for large system sizes $L\ge 768$ are
best described by the single logarithmic form, $C= a \ln\left(bL\right)$ expected at the clean Ising critical point.

How can we explain these observations?
In the present system, both the bare disorder strength and the coupling $\epsilon$
are rather weak. The renormalization group flow (Fig.\ \ref{fig:Cardy_flow}) therefore does not travel
too far from the origin, explaining that our effective exponents are very close
to the clean Ising ones. Note, however, that the renormalization group ``time'' (flow parameter) and, correspondingly,
the system size range needed to go around the loop in Fig.\ \ref{fig:Cardy_flow}
do not vary much with the size of the loop. As the bare disorder is weak, the system thus does
not reach the falling, asymptotic part of the loop even for $L=1600$. This may also explain why
the effective correlation length exponent $\nu_{\rm eff}=0.93(2)$ that we extract from the slope of
the Binder cumulant vs.\ temperature curves in this weakly disordered system
does not fulfill the Chayes' inequality\cite{CCFS86} $d\nu \ge 2$.
Because of the exponential dependence of the breakup length $L_b$ on the disorder strength,
confirming this picture numerically would likely require enormous system sizes.

%%%%%%%%%%%%%%%%%%%%%%%%%%%%%%%%%%%%%%%%%%%%%%%%%%%%%%%%%%%%%%%%%%%%%%%%%%%%%%%%%
\section{Conclusions}
\label{Conclusions}
%%%%%%%%%%%%%%%%%%%%%%%%%%%%%%%%%%%%%%%%%%%%%%%%%%%%%%%%%%%%%%%%%%%%%%%%%%%%%%%%%

In summary, we have performed high-accuracy Monte Carlo simulations of the disordered
three-color Ashkin-Teller model in two dimensions using systems with up to $1600^2$
lattice sites. We have investigated two types of
disorder, random site dilution and random interactions (bond randomness).
Our results show that the first-order phase transition of the clean Ashkin-Teller model
is destroyed by the randomness, in agreement with the Aizenman-Wehr
theorem.\cite{ImryMa75,*ImryWortis79,*HuiBerker89,AizenmanWehr89}

We have carefully analyzed the critical behavior of the emerging continuous phase
transition and found strong evidence that the asymptotic critical behavior is
universal and agrees with the predictions of Cardy's renormalization group theory.
\cite{Cardy96,*Cardy99} This means, the critical exponents coincide with those of the
clean two-dimensional Ising model, but with additional logarithmic corrections to scaling
analogous to those found in the disordered two-dimensional Ising model. For example,
the specific heat takes the characteristic double-logarithmic form (\ref{eq:C_lnln}).

What could be the reason for the differences between our results and the unusual behavior
(nonuniversal critical exponents and violations of the inequality $d\nu\ge 2$ due to Chayes et al.\cite{CCFS86})
reported in Refs.\ \onlinecite{BellafardKatzgraberTroyerChakravarty12,BellafardChakravartyTroyerKatzgraber14}?
First, our systems are significantly larger: Refs.\ \onlinecite{BellafardKatzgraberTroyerChakravarty12}
and \onlinecite{BellafardChakravartyTroyerKatzgraber14} used systems with up to $32^2$
and $128^2$ sites, respectively, while our systems have up to $1600^2$ sites. As the
Ashkin-Teller model crosses over very slowly from the first-order transition of the clean problem to the
continuous transition of the disordered one, simulations of smaller
systems are, perhaps, not sufficient to reach the asymptotic regime.
Large systems are particularly important if the disorder strength is small. In fact, our own simulations for
the weak dilution $p=0.1$ show that, depending on $\epsilon$, the asymptotic regime may only be reached
for $L \ge 400$. The random-bond system with $J_h=6/5$ and $J_l=4/5$ is especially weakly
disordered; correspondingly, it does not reach the asymptotic regime even for $L=1600$.

This interplay between the disorder strength and the cross-over between first-order
and continuous transitions is also borne out by the analysis of the correlation length exponent $\nu$.
The asymptotic finite-size scaling (\ref{eq:dRdT_ln}) of dimensionless quantities such as the Binder cumulant
leads to an effective exponent $\nu_{\rm eff} >1$. This is what we have observed in all our systems
except for the one with the weakest disorder, viz., the random-bond system with with $J_h=6/5$ and $J_l=4/5$.
This supports the notion that the correlation length exponents reported
in Refs.\ \onlinecite{BellafardKatzgraberTroyerChakravarty12,BellafardChakravartyTroyerKatzgraber14}
may be effective exponents outside the asymptotic regime.

We note, however, that a significant discrepancy between our data and those reported in
Ref.\ \onlinecite{BellafardKatzgraberTroyerChakravarty12}
is manifest already in the clean phase diagram, Fig.\
\ref{fig:pd_clean}, where system size effects should be less important.
 Our phase boundary agrees with the old data by Grest and Widom \cite{GrestWidom81}
but disagrees with Ref.\ \onlinecite{BellafardKatzgraberTroyerChakravarty12}.

As a byproduct, our simulations for $\epsilon=0$ (where the Ashkin-Teller Hamiltonian is equivalent to three independent
Ising models) also help to resolve the long-standing controversy about the critical behavior of the
disordered two-dimensional Ising model. Our large-scale data for systems with up to $2240^2$ sites
provide strong support for the logarithmic-correction (strong-universality)
scenario \cite{DotsenkoDotsenko83,Shalaev84,Shankar87,*Shankar88,Ludwig88} according to which the
critical behavior is characterized by the clean Ising exponents and universal logarithmic corrections.

We now put our results in the general context of phase transitions of two-dimensional disordered systems.
Following the analytical results on the disordered two-dimensional Ising \cite{DotsenkoDotsenko83,Shalaev84,Shankar87,*Shankar88,Ludwig88}
and Ashkin Teller \cite{Cardy96} models, it was conjectured that all critical behavior in two-dimensional disordered systems
belongs to the disordered Ising universality class. This belief in super-universal critical behavior was further strengthened by
early numerical results for disordered Ising,\cite{ADSW90,WSDA90a,WSDA90b} Ashkin-Teller,\cite{WisemanDomany95}
and Potts\cite{WisemanDomany95,ChenFerrenbergLandau95} models as well as heuristic interface arguments.
\cite{KSSD95}
However, later simulations of the disordered $q$-state Potts model\cite{JacobsenCardy98,ChatelainBerche98}
belied these expectations: They showed that
the exponent $\beta/\nu$ does depend on the value of $q$ and generally differs from the Ising value of 1/8.
Recently, unexpectedly complex behavior was also found in the two-dimensional random-bond Blume-Capel model,
\cite{MBHF09,MBHFP10,TheodorakisPanagiotisFytas12} an Ising-like spin-1 model with an additional single-ion
anisotropy.

Cardy's renormalization group approach \cite{Cardy96,*Cardy99} was generalized by Pujol \cite{Pujol96}
from $N$ coupled Ising models to $N$ coupled $q$-state Potts models. For $q=2$ (the Ising case), Pujol's
results agree with Cardy's. For $q>2$, however, he found the emerging critical behavior
to be controlled by a nontrivial random fixed point. Testing these predictions numerically remains a task
for the future.

Finally, the quantum version of the Ashkin-Teller model has recently attracted considerable attention
in connection with the question of how first-order \emph{quantum} phase transitions react to disorder.
Strong-disorder renormalization group calculations predict infinite-randomness critical points
in different universality classes, depending on the coupling strength $\epsilon$.
\cite{GoswamiSchwabChakravarty08,HrahshehHoyosVojta12,Barghathietal14}
Moreover, the two-color model is predicted to feature an unusual strong-disorder infinite-coupling
phase \cite{HHNV14}. Our Monte Carlo method can be easily generalized from the two-dimensional
classical case to the $(1+1)$-dimensional quantum case. Some calculations along these lines are under
way.

%%%%%%%%%%%%%%%%%%%%%%%%%%%%%%%%%%%%%%%%%%%%%%%%%%%%%%%%%%%%%%%%%%%%%%%%%%%%%%%%%
\section*{Acknowledgements}
%%%%%%%%%%%%%%%%%%%%%%%%%%%%%%%%%%%%%%%%%%%%%%%%%%%%%%%%%%%%%%%%%%%%%%%%%%%%%%%%%
This work was supported by the NSF under Grants No.\ DMR-1205803
and No. PHYS-1066293, by Simons Foundation, by FAPESP under Grant No.\ 2013/09850-7, by CNPq under Grants
No.\ 590093/2011-8 and No.\ 305261/2012-6, by the 973 Program under Project No.\ 2012CB927404,
and by the NSFC under Grant No.\ 11174246.
 J.H. and T.V. acknowledge the hospitality of the
Aspen Center for Physics. R.N. thanks the High Performance Computing Facility at IIT-Madras
for computational assistance and resources.

%%%%%%%%%%%%%%%%%%%%%%%%%%%%%%%%%%%%%%%%%%%%%%%%%%%%%%%%%%%%%%%%%%%%%%%%%%%%%%%%%
\appendix
%%%%%%%%%%%%%%%%%%%%%%%%%%%%%%%%%%%%%%%%%%%%%%%%%%%%%%%%%%%%%%%%%%%%%%%%%%%%%%%%%
\section{Short Monte Carlo runs and unbiased estimators}
\label{sec:bias}
%%%%%%%%%%%%%%%%%%%%%%%%%%%%%%%%%%%%%%%%%%%%%%%%%%%%%%%%%%%%%%%%%%%%%%%%%%%%%%%%%

Short Monte Carlo runs consisting of only a small number of measurements per sample
introduce biases into some observables, at least if one employs the usual estimators.
Consider, for example, the magnetic susceptibility (of a single sample) which is related
to the variance of the magnetization via
$\chi = (L^2/T) \sigma_M^2$ with $\sigma_M^2 = \langle m^2 \rangle -\langle m \rangle^2$.
In a Monte Carlo simulation, the variance $\sigma_M^2$ is usually replaced by the
estimator
\begin{equation}
s_M^2 = \frac {1}{n_m} \sum_{i=1}^{n_m} m_i^2 - \left ( \frac {1}{n_m} \sum_{i=1}^{n_m} m_i \right)^2
\label{eq:chi_estimator_wrong}
\end{equation}
where $m_i$ is the magnetization of an individual measurement and $n_m$ is their number.
It is well known in statistics that $s_M^2$ underestimates the variance, even for uncorrelated $m_i$.
This can be seen by evaluating the expectation value of  $s_M^2$ as
\begin{eqnarray}
\langle s_M^2 \rangle &=& \frac {1}{n_m} \sum_{i=1}^{n_m} \langle m_i^2 \rangle - \frac {1}{n_m^2} \sum_{i=1}^{n_m} \sum_{j=1}^{n_m} \langle m_i m_j \rangle \nonumber\\
&=& \langle m^2 \rangle - \frac {1}{n_m} \langle m^2 \rangle - \frac {n_m-1}{n_m} \langle m \rangle^2\nonumber \\
&=& \sigma_M^2 \left(1 -  \frac {1}{n_m} \right)~.
\label{eq:<s^2>}
\end{eqnarray}
If the $m_i$ are correlated with a correlation time of $\tau$, the bias becomes even stronger:
$\langle s_M^2 \rangle \approx \sigma_M^2 \left(1 -   ({1+A})/{n_m} \right)$ with $A \sim \tau$.
Other quantities defined as variances or covariances develop analogous biases, including
the specific heat
$C = (L^2/T^2) ( \langle e^2 \rangle -\langle e \rangle^2) $.

Note that these biases are not important in normal Monte Carlo simulations that consist of
one (long) run  of $n_m$ measurements because the bias decays as $n_m^{-1}$ while the statistical
error decays as $n_m^{-1/2}$. The bias is thus much smaller than the statistical error and
can be neglected. However, if the results of short runs are averaged over a large number $n_s$
of samples, this argument changes. The bias still decays as $n_m^{-1}$ but the statistical
error and sample-to-sample fluctuations due to disorder are  suppressed by an additional factor $n_s^{-1/2}$.
It is thus clear that the bias cannot be neglected for a sufficiently large number of samples.
(If the disorder-induced sample-to-sample fluctuations are weaker than the thermodynamic fluctuations,
this is expected when $n_s \gtrsim n_m$. In the opposite case, for strong disorder fluctuations,
the bias becomes important roughly when $n_s \gtrsim n_m^2$.)

How can one correct the bias due to short Monte Carlo runs? If the measurements were completely
independent, one could simply multiply the usual estimator (\ref{eq:chi_estimator_wrong}) by $n_m/(n_m-1)$.
However, achieving full independence requires long time intervals between consecutive measurements
which makes the simulations inefficient. We instead introduce modified, unbiased estimators.
To this end, we split the Monte Carlo run of $n_m$ measurements into two halfs, each with
$n_m/2$ measurements. We also perform a few extra Monte Carlo sweeps between the two halfs to ensure
that they are independent of each other. The improved estimator of $\sigma^2_M$ is then given by
\begin{equation}
\tilde s_M^2 = \frac {1}{n_m} \sum_{i=1}^{n_m} m_i^2 - \left [ \frac {2}{n_m} \sum_{i=1}^{n_m/2} m_i \right] \left [ \frac {2}{n_m} \sum_{i=n_m/2+1}^{n_m} m_i \right].
\label{eq:chi_estimator_unbiased}
\end{equation}
Following the same steps as outlined in (\ref{eq:<s^2>}), it is straightforward to show that
$\langle \tilde s_M^2 \rangle = \sigma_M^2$. This means $\tilde s_M^2$ is unbiased.
An analogous unbiased estimator can be defined for the specific heat $C$.

In Sec.\ \ref{subsec:Observables}, we defined two magnetic Binder cumulants, $g_\textrm{av}$
and $g_\textrm{gl}$, as well as two correlation lengths, $\xi_\textrm{av}$ and $\xi_\textrm{gl}$.
The ``average'' versions $g_\textrm{av}$ and $\xi_\textrm{av}$ suffer from short-run biases similar
to those discussed above while the ``global'' versions $g_\textrm{gl}$ and $\xi_\textrm{gl}$
are unbiased. In principle, one could correct the biases in $g_\textrm{av}$ and $\xi_\textrm{av}$
by using improved estimators. However these would have a more complicated structure than
(\ref{eq:chi_estimator_unbiased}) to deal with the terms in the denominators of eqs.\ (\ref{eq:Binder})
and (\ref{eq:xi_av}).
For simplicity, we have not done this. Instead we mostly rely on the unbiased observables
$g_\textrm{gl}$ and $\xi_\textrm{gl}$. (Interestingly, our numerical data suggest that
$g_\textrm{av}$ and $\xi_\textrm{av}$ have significantly smaller biases than $C$ and $\chi$.)

\bibliographystyle{apsrev4-1}
\bibliography{../00bibtex/rareregions}

\end{document}